\documentclass[conference]{IEEEtran}
\usepackage[explicit]{titlesec}
\usepackage{graphicx} 
\usepackage{multirow}
\usepackage{booktabs}
\usepackage{easyReview} 
\usepackage{verbatim} 
\usepackage{epstopdf} 
\usepackage{setspace} 
\usepackage{mathtools} 
\usepackage{mdframed} 
\usepackage{subfigure} 
\usepackage{algorithmic}
\usepackage{amsmath}
\usepackage{amsthm}
\usepackage{mathrsfs}
\usepackage{extarrows}
\usepackage{cite}

\newtheorem{theorem}{Theorem}

\usepackage{bbm}
\usepackage[bookmarks=false, colorlinks=true, citecolor=blue]{hyperref}
\usepackage{cleveref}
\usepackage{amssymb}
\newtheorem{remark}{Remark}

\newcommand{\RNum}[1]{\uppercase\expandafter{\romannumeral #1\relax}}
\newtheorem{Lemma}{Lemma}

\newtheorem{problem}{Problem}

\usepackage{stfloats}
\usepackage{color, soul}
\usepackage{booktabs}
\usepackage{tikz,xcolor}
\usetikzlibrary{arrows.meta,automata, positioning,,backgrounds,matrix}

\newenvironment{proofsketch}{\begin{proof}[\textit{Proof Sketch}]}{\end{proof}}
\definecolor{lime}{HTML}{A6CE39}
\usepackage[left=0.68in,right=0.68in,top=0.67in,bottom=0.99in]{geometry}
\usepackage{xr}

\titlespacing{\section}{0pt}{1.2ex plus .0ex minus .0ex}{.3ex plus .0ex}
\titlespacing{\subsection}{0pt}{1.2ex plus .0ex minus .0ex}{.3ex plus .0ex}

\makeatletter
\foreach \x in {A, ..., Z}{%
	\expandafter\xdef\csname orcid\x\endcsname{\noexpand\href{https://orcid.org/\csname orcidauthor\x\endcsname}{\noexpand\orcidicon}}
}

\newcommand*\bigcdot{\mathpalette\bigcdot@{.5}}
\newcommand*\bigcdot@[2]{\mathbin{\vcenter{\hbox{\scalebox{#2}{$\m@th#1\bullet$}}}}}
\makeatother

\usepackage[linesnumbered,ruled,vlined]{algorithm2e}
	
	\IEEEoverridecommandlockouts
	\makeatletter
	\def\@thmnote#1{\textit{[#1]}} 
	\makeatother

\begin{document}
\title{Age-Optimal Sampling and Routing under Intermittent Links and Energy Constraints}
\author{
			A. Utku Atasayar, Aimin Li, \emph{Member, IEEE}, Çağrı Arı, and 
			Elif Uysal, \emph{Fellow, IEEE}\\
		 \\
	\vspace{-3em}
    \thanks{An earlier version of this work was presented in part at the ACM International Symposium on Mobile Ad Hoc Networking and Computing (ACM MobiHoc) 2025 \cite{Mobihoc25}. A. Utku Atasayar, Aimin Li, Çağrı Arı, and Elif Uysal are with \href{https://cng-eee.metu.edu.tr/}{Communication Networks Research Group (CNG)}, EE Dept, Middle East Technical University (METU), Ankara, 06800, Turkiye (e-mail: atasayar.utku@metu.edu.tr;  aimin@metu.edu.tr;  ari.cagri@metu.edu.tr;  uelif@metu.edu.tr). This work is supported by the European Union through ERC Advanced Grant 101122990-GO SPACE-ERC-2023-A and The Scientific and Technological Research Council of Türkiye (TÜBİTAK) 1515 Frontier R\&D Laboratories Support Program for Türk Telekom 6G R\&D Lab under project number 5249902. Views and opinions expressed are however those of the author(s) only and do not necessarily reflect those of the European Union or the European Research Council Executive Agency. Neither the European Union nor the granting authority can be held responsible for them. (\textit{Corresponding Authors: Aimin Li and Elif Uysal}). Code is available at \url{https://github.com/auatasayar/Joint_Sampling_and_Routing_for_AoI_Minimization}.
    }
    }
		\maketitle
		\allowdisplaybreaks

\begin{abstract}
Links in practical systems, such as satellite--terrestrial integrated networks, exhibit distinct delay distributions, intermittent availability, and heterogeneous energy costs. These characteristics pose significant challenges to maintaining timely and energy-efficient status updates. While link availability restricts feasible transmission routes, routing decisions determine the actual delay and energy expenditure. This paper tackles these challenges by jointly optimizing sampling and routing decisions to minimize monotonic, non-linear Age of Information (AoI). The proposed formulation incorporates key system features, including multiple routes with correlated random delays, stochastic link availability, and route-dependent energy consumption. We model the problem as an infinite-horizon Constrained Semi-Markov Decision Process (CSMDP) with a hybrid state--action space and develop an efficient nested algorithm, termed Bisec-\textsc{ReaVI}, to solve this problem. We analyze the structural properties of the solution and reveal a well-defined jointly optimal policy structure: (i) For general monotonic penalty functions, the optimal sampling policy is a piecewise linear waiting policy with at most $N$ breakpoints given $N$ routes; and (ii) under a derived Expected Penalty Ordering condition, the optimal routing policy is a monotonic threshold-based handover policy characterized by at most $\binom{N}{2}$ thresholds. Numerical experiments in a \textit{satellite--terrestrial} integrated routing scenario demonstrate that the proposed scheme efficiently balances energy usage and information freshness, and reveal a counter-intuitive insight: \textit{even routes with higher average delay, higher delay variance or lower availability can still play a critical role in minimizing monotonic functions of AoI}.
\end{abstract}
        \IEEEpeerreviewmaketitle
\section{Introduction}
\subsection{Background}

In an increasingly connected world where systems rely on remotely sampled data to make real-time decisions, the freshness of data samples has become a key driver of application performance. Hence, information freshness is emerging as a Key Performance Indicator (KPI) across a wide range of applications, supported by next-generation communication networks spanning wired, wireless, and non-terrestrial links. For instance, in remote-sensing-based emergency response systems, access to \textit{fresh} data regarding environmental variables supports real-time risk assessment and enhances response efficiency. Similarly, in Vehicle to Everything (V2X) scenarios, vehicles rely on continuously updated information collected through multi-sensor data fusion to navigate safely and adapt to rapidly changing environmental conditions. Moreover, in Industrial Internet of Things (IIoT) applications, the staleness of sensor data negatively impacts production efficiency, equipment maintenance timing, and overall operational safety.

This growing emphasis on information freshness has led to the development of the Age of Information (AoI) metric that quantifies it~\cite{kaul2012real}. Distinct from traditional latency, AoI provides a receiver-centric, flow-level measure of information timeliness. Formally, at any time $t$, the AoI is defined as $\Delta(t) \triangleq t - U(t)$, where $U(t)$ denotes the generation time of the latest received sample. Maintaining a low AoI requires both sufficiently frequent updates and low-latency delivery, thus coupling throughput and delay in a novel performance metric. Hence, AoI combines the conventional performance metrics of latency and throughput in a novel way. Over the past few years, AoI minimization has been studied under various constraints and network models, including energy-limited update policies~\cite{bacinoglu2015age, yates2015lazy, bacinoglu2019optimal}, multi-hop and multi-source networks~\cite{talak2017minimizing, beytur2018minimizing, 11269014, kadota2019scheduling, bedewy2021optimal,  li2025optimal,ceran2021reinforcement}, broadcast settings~\cite{kadota2018scheduling}, and unreliable or error-prone communication channels~\cite{7541765, 9931172,11184587} (see \cite{DBLP:journals/jsac/YatesSBKMU21a} for a comprehensive review). Furthermore, in scenarios where the significance of data extends beyond temporal freshness, more sophisticated evaluation frameworks have been developed~\cite{chen2022uncertainty, chen2024optimal, sun2019sampling, 10806969, 10807024, shisher2024timely, sun2019sampling2}. Some of these frameworks utilize AoI as an intermediate metric to capture task-specific relevance through the freshness of data samples~\cite{shisher2024timely}.   

\subsection{Motivation}
In this paper, we focus on the optimization of monotonic functions of AoI. Our goal is to extend the formulation of Age-optimal sampling first proposed in \cite{sun2017update} to a case where there are multiple routing options between the source and the destination. The new formulation proposed in this paper was inspired and motivated by the growing interest in Satellite IoT and integrated TN-NTN in 5G and 6G, where data transmission decisions are faced with choosing between routing through non-terrestrial links versus terrestrial connections:
\begin{itemize}
	\item[(i)] Routing through \textbf{terrestrial links} typically offers low-latency and energy-efficient transmission, owing to the relatively short propagation distance and mature ground-based infrastructure such as optical fiber and cellular networks. These links are generally stable under normal operating conditions and can support high-throughput, delay-sensitive services. However, their performance and reliability are susceptible to \textit{network congestion}, \textit{coverage holes}, and \textit{infrastructure failures.}
	\item[(ii)] Routing through \textbf{non-terrestrial links} often exhibits intermittent availability because the space–atmosphere channel and satellite geometry change over time. In addition to orbital dynamics and visibility windows, beam/footprint handovers, gateway reassociation, and ISL route reconfiguration introduce short disruptions that fragment service into available/unavailable epochs. Propagation conditions vary with rain attenuation (Ku/Ka), cloud ice/water content, ionospheric irregularities, and geomagnetic storms, which can drive rapid SNR fluctuations and temporary outages. Even during physical available periods, random-access collisions, beam scheduling, and backhaul bottlenecks can make access bursty.
\end{itemize}
	
Moreover, delay statistics across routes may be \textit{correlated} due to shared infrastructures, spectrum usage, or satellite visibility patterns, which further couples the sampling and routing decisions. These practical considerations result in route-dependent delay statistics, stochastic link availability, and heterogeneous energy costs, which motivates a unified framework that jointly optimizes both the sampling time and the routing path. Our formulation addresses the fundamental question of how route selection impacts information freshness in hybrid terrestrial/non-terrestrial environments, and provides a theoretical foundation for age-aware joint sampling and routing in next-generation communication networks.

\subsection{Related Works and Contributions}
\begin{figure}[t]
    \centering
    \includegraphics[width=0.9\linewidth]{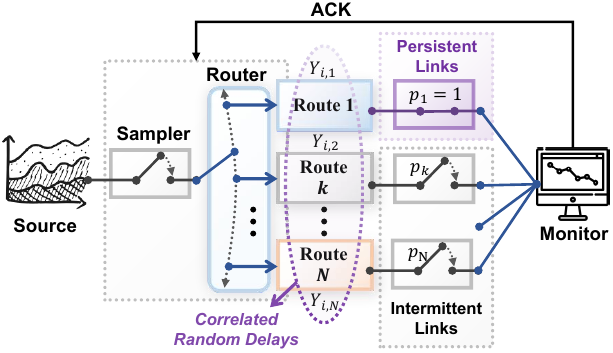} 
    \caption{A remote monitoring system, where status updates are transmitted through $N$ heterogeneous routes.}
    \label{sysmodel}
\end{figure}
\begin{itemize}
    \item \textbf{System Model}: We formulate a joint sampling and routing problem in which a transmitter optimizes both the \textit{sampling interval} and the \textit{transmission route} to minimize the \textit{long-term average of a monotonic function of the AoI} at the destination, subject to an average routing energy constraint and route availability constraint. Our formulation is a direct extension of the problem in \cite{sun2017update}, and in contrast to this and other prior work where the sampling problem is attacked under given delay statistics \cite{sun2019sampling,10004990,10806969,10807024}, our system model takes a \textit{proactive} approach by actively selecting and switching routes to control the delay experienced by status updates. Meanwhile, unlike existing multi-channel scheduling problems that either focus on homogeneous channels with uniform one-slot delays \cite{sun2023age} or heterogeneous channels where each channel experiences an on-off constant discrete delay \cite{pan2021minimizing,9882370}, our work considers distinct \textit{continuous delay distributions} across different routes and jointly optimizes sampling and route selection strategies. This work generalizes our prior work in \cite{Mobihoc25} by incorporating the following practical aspects: (i) \textit{Correlated route delays}, motivated by shared infrastructures or congestion effects that couple delay statistics across different routes; (ii) \textit{Stochastic route availability}, modeled as independent random on/off links with fixed availability probabilities, capturing real-world uncertainties such as satellite visibility, atmospheric attenuation, and space-weather-induced disruptions; (iii) \textit{Route-dependent energy consumption} under an average energy constraint, which is essential for energy-limited platforms such as satellites or IoT devices; and (iv) \textit{Non-linear age penalty functions}, extending the optimization framework beyond standard linear AoI to support general monotonic functions of age, allowing for more flexible service requirement modeling. 
    \item \textbf{Solution Methodology}: We show the problem can be formulated by a Constrained Semi-Markov Decision Process (CSMDP) with \textit{uncountable hybrid} state and action spaces. The state captures discrete link availability and continuous delay, while the action includes both routing choices and sampling intervals. Such CSMDPs are known to be challenging to solve due to the size of the state and action spaces, and previous research has addressed this complexity by: ($i$) discretizing the \textit{uncountable} state space and the action space \cite{8712546,zhou2024age}, which introduces \textit{quantization error} or; ($ii$) focusing on a special case of SMDP where the state transitions are independent of actions\footnote{This allows the Markov decision process to be reduced to a renewal reward process, thus simplifying the analysis.} \cite{sun2019sampling,sun2017update,10004990}, which, however, does not hold in our scenario. In this paper, we develop a new nested algorithm namely Bisection \textit{\textbf{R}elative \textbf{E}xpected \textbf{A}ction \textbf{V}alue \textbf{I}teration} (\textsc{Bisec-ReaVI}) that directly solves this class of SMDPs without discretizing the space. \textit{To the best of our knowledge, this is the first algorithm that efficiently solves hybrid-state CSMDPs while preserving structural optimality and avoiding discretization error.}
    \item \textbf{Structural Results}: We prove that each of the jointly optimal sampling and routing policies exhibit a graceful \textit{threshold structure}: ($i$) The routing policy is a thresholds-based handover policy, where a specific route is selected when the current AoI at the receiver falls within certain range (a route may be optimal for more than one such regions), precisely determined by multiple thresholds; ($ii$) A new sample is taken and transmitted when the AoI at the receiver reaches a threshold that is a function of the selected transmission route. These structural properties deem the policies suitable for practical implementation.
    \item \textbf{Counter-Intuitive Insights:} We test our algorithms on the model of an \textit{integrated satellite-terrestrial} communication network scenario. Our simulation results reveal an intriguing insight: \textit{routes with higher mean delay, greater variance, or lower availability can still contribute to minimizing AoI.} This finding challenges conventional wisdom that may prioritize routes with minimal mean delay {or delay variance} characteristics. It demonstrates that the strategic utilization of diverse routing options can lead to superior information freshness.
\end{itemize}

\section{System Model}

We consider a remote monitoring system, as illustrated in Fig.~\ref{sysmodel}, consisting of a source, a sampler, a router, and a monitor. Status updates are timely generated, and each is transmitted through one of the $N$ heterogeneous routes, with the objective of maintaining the freshest possible information on the monitor at all times.
\subsection{\textbf{Persistent and Intermittent Links}}
In this work, we consider a heterogeneous network consisting of multiple communication routes, denoted collectively as the set $\mathcal{N}$. These communication routes are categorized into two disjoint subsets based on their long-term availability:

\begin{itemize}
    \item \textbf{Persistent Routes} ($\mathcal{R}_{\infty}$): The routes in set $\mathcal{R}_{\infty}$ are continuously accessible over time and typically correspond to terrestrial links such as fiber-optic or cellular infrastructure. Due to their stable physical environment and minimal susceptibility to external disruptions, these links exhibit deterministic availability, and are modeled with an availability probability of $p_k = 1$. The set of persistent routs are given as: 
    $
        \mathcal{R}_{\infty}=\{k\in\mathcal{R}:p_k=1\}\subset\mathcal{N}.
   $ 
    To guarantee continuous data transmission, we assume that $\mathcal{R}_{\infty} \neq \emptyset$, which corresponds to the presence of at least one persistently available terrestrial route (e.g., cellular or fiber-optic), ensuring baseline connectivity even when all intermittent routes are unavailable.

    \item \textbf{Intermittent Routes} ($\mathcal{R}_{<\infty}$): This subset includes routes whose availability varies over time due to stochastic physical factors. Typical examples include satellite links or other opportunistic channels affected by satellite orbital motion, line-of-sight (LOS) constraints, or environmental interference (e.g., weather conditions). Each intermittent route $k \in \mathcal{R}_{<\infty}$ is characterized by a stationary availability probability $p_k \in (0,1)$, which denotes the long-term fraction of time the route is usable. The set of persistent routs are given as: 
 ${R}_{<\infty}=\{k\in\mathcal{R}:0<p_k<1\}\subset\mathcal{N}.$
\end{itemize} 

\subsection{Correlated Random Delays}
The transmission delays across different routes at a given transmission instance may exhibit \textit{statistical dependence}. Let \( Y_{i,k} \) denote the random transmission delay experienced when the \(i\)-th packet is sent via route \(k \in \mathcal{N}\). We model the delay vector at time \(i\) as \( \mathbf{Y}_i \triangleq (Y_{i,1}, Y_{i,2}, \ldots, Y_{i,N}) \), which is assumed to be drawn from a stationary multivariate distribution \( \mathbf{Q} \). This distribution captures both the variability and potential correlations among the different routes at each transmission instance. Let \( F_{\mathbf{Y}}(y_1, \ldots, y_N) \) denote the joint cumulative distribution function (CDF) of \( \mathbf{Y}_i \). This model reflects realistic phenomena such as correlated queuing delays, weather-induced impairments, or congestion that may simultaneously affect multiple communication routes.

We denote the marginal distribution of route \(k\) by \(Q_k\), and assume that the per-transmission delays \(\{Y_{i,k}\}_{i \in \mathbb{N}^+}\) are i.i.d. over time with \(Y_{i,k} \sim Q_k\) for each \(k \in \mathcal{N}\). The mean and variance of the delay on route \(k\) are denoted by: 
\begin{equation}
    \mu_k \triangleq \mathbb{E}[Y_{i,k}] < \infty, \quad 
\sigma_k^2 \triangleq \mathbb{E}[(Y_{i,k} - \mu_k)^2] < \infty.
\end{equation}
\begin{remark}[Sufficiency of Marginals]
    Although the delay vectors $\mathbf{Y}_i$ may exhibit correlation across routes within the same epoch (e.g., due to shared weather events affecting multiple satellite links), the optimal control policy depends exclusively on the marginal distributions $Q_k$. 
Since the system state resets upon delivery (due to the i.i.d. nature of epochs) and the decision maker observes only the realization of the selected route, the joint correlation structure does not provide exploitable information for future epochs. Thus, the optimization problem reduces to one defined purely by the set of marginals $\{Q_k\}_{k=1}^N$, and the potential correlation between unselected routes does not impact the achievable optimal age.
\end{remark}

In the subsequent analysis, where the specific route index is determined by the policy, we will omit the subscript \(k\) and denote the delay incurred during the \(i\)-th epoch simply as \(Y_i\).

\subsection{Heterogeneous Energy Costs}
The system is subject to a long-term average energy constraint denoted by \( E_{\max} \), which limits the energy consumption over time. This constraint is particularly relevant in \textit{energy-constrained systems} such as remote sensing applications or satellite-terrestrial networks, where power sources (e.g., battery-powered ground terminals or solar-powered satellite relays) are limited. There are two types of energy expenditures:

\begin{itemize}
    \item \textbf{Sampling Energy Cost:} Every time a new status update is generated, a fixed amount of energy \(C_s > 0\) is incurred. This cost accounts for sensing, computation, and other acquisition overheads required to produce a fresh update.

    \item \textbf{Transmission Energy Cost:}
    Upon sampling the $i$-th packet, the generated update is transmitted through a selected route \(R_{i-1}\), which incurs a \textit{per-unit-time transmission energy cost}, denoted by $E(t)$, given as
    \begin{equation}
        E(t)=\begin{cases}
           G_{R_{i-1}}, & t\in(S_i,D_i]\\
           0, & t\in(D_{i-1},S_i]
        \end{cases},
    \end{equation}
    where $G_{k}$ is the per-unit-time transmission cost when the route $k$ is busy. This cost reflects factors such as propagation loss, transmission power requirements, protocol configurations, and hardware-level energy consumption. 
\end{itemize}

This energy model introduces a trade-off between \textit{timeliness} (i.e., age of information) and \textit{energy efficiency}. For example, lower-delay routes may be intermittently available but may also incur higher energy per unit time (e.g., high-bandwidth satellite links), whereas persistent links might offer lower energy efficiency due to higher latency.

We assume a \textit{non-preemptive} system, where a new transmission can begin only after the previous one has been completed \cite{sun2017update}. Upon receiving each data sample, the monitor sends an ideal acknowledgment (ACK) to the transmitter, indicating that the system is ready to initiate the next transmission.

We adopt the \textit{generate-at-will} model~\cite{sun2017update, yates2015lazy}, in which the sampler can become active at any time, provided that a new transmission is allowed. We next introduce some notation. After receiving the ACK corresponding to the $i$-th transmission, the $(i+1)$-th data sample is generated and submitted to route $R_i$, selected from the pool of available routes $\mathcal{R}_i$, at time instant $S_{i+1}$. It is subsequently delivered to the monitor at time instant ${D_{i+1} = S_{i+1} + Y_{i+1}}$. The overall energy cost for transmitting the \((i+1)\)-th update is:
\begin{equation}\label{eq.5}
    E_{i+1} = C_s + G_{R_i} Y_{i+1}.
\end{equation}

\subsection{Age of Information and No-linear Age}
\begin{figure}[t]
    \centering
    \includegraphics[width=0.85\linewidth]{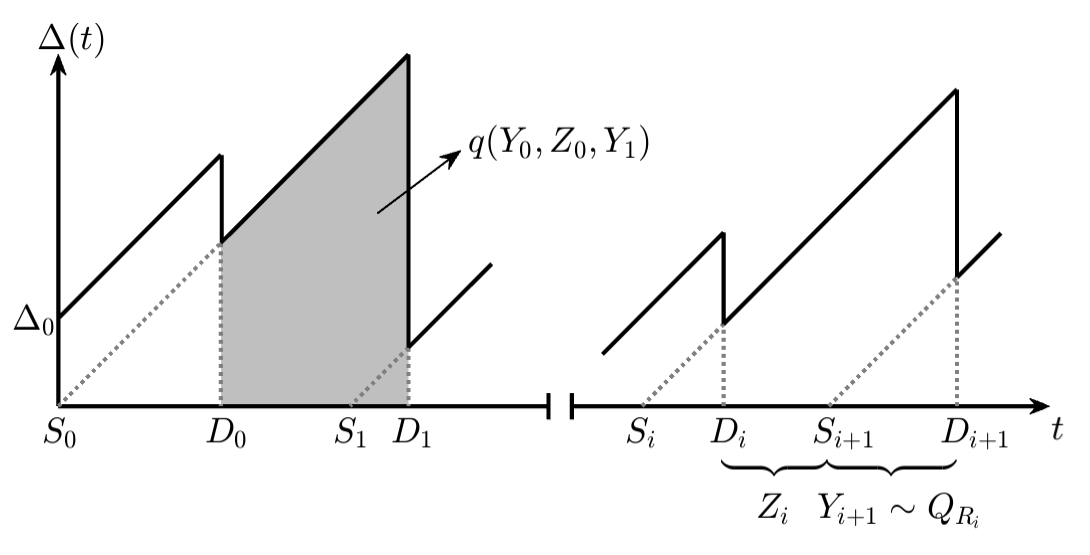}
    \caption{Sample evolution of the AoI process $\Delta(t)$.}
    \label{fig_age}
    \vspace{-0.21 in}
\end{figure}

The Age of Information (AoI) is the metric of our interest to measure the \textit{freshness} of information. This metric is defined as the time elapsed since the generation of the most recently received data sample~\cite{kaul2012real}. Specifically, the AoI $\Delta(t)$ at time $t$ is defined by
\begin{equation}\label{aoidefinition}
\Delta(t) \triangleq t - S_i, \quad \text{if  } D_i \leq t < D_{i+1}.
\end{equation} 
The AoI $\Delta(t)$ is a stochastic process that increases linearly over time and experiences downward jumps to the most recent delay value $Y_i$ upon the delivery of the $i$-th data sample at time $D_i$, as illustrated in Fig.~\ref{fig_age}. The value of $\Delta(t)$ between the time instants $S_0 = 0$ and $D_0$ is assumed to increase linearly, starting from an arbitrary finite initial real value $\Delta(0) = \Delta_0 < \infty$.

While the standard linear AoI metric captures the timeliness of updates, many practical systems exhibit sensitivity to information staleness that is non-linear. To address this, we consider a generalized \textit{age penalty function} $f(\Delta(t))$, where $f(\cdot): [0, \infty) \to [0, \infty)$ is a continuous, monotonic non-decreasing function. This formulation allows us to model diverse application requirements: the standard linear case corresponds to $f(\Delta) = \Delta$, while scenarios demanding stricter freshness constraints can be modeled using exponential penalties and applications with saturation effects can be captured by logarithmic functions. 

Furthermore, to ensure the problem is well-posed and the expected costs are finite, we adopt the following assumption on the penalty function. For any route $k \in \mathcal{N}$, the delay distribution $Q_k$ satisfies the finite expected penalty condition: $\mathbb{E}_{Y \sim Q_k} [f(Y)] < \infty.$

\section{Problem Formulation}\label{section2.4}

\begin{figure}[t]
    \centering
    \includegraphics[width=0.85\linewidth]{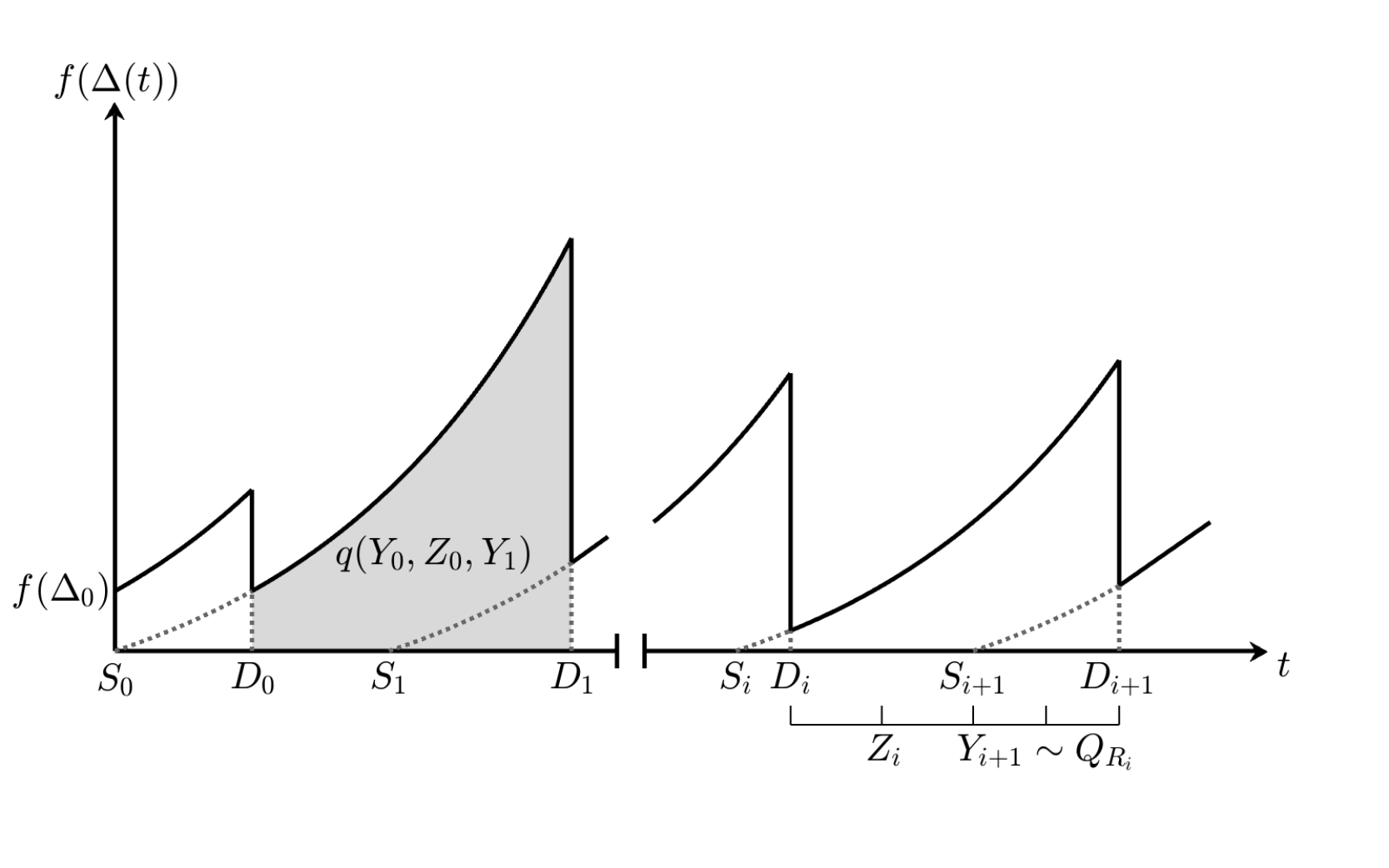}
    \caption{Sample path of the non-linear $f(\Delta(t)) = e^{0.35 \Delta(t)} - 1$.}
    \label{fig_age_exp}
    \vspace{-0.21 in}
\end{figure}

We aim at minimizing the long-term average penalty of AoI by designing a joint sampling and routing policy $\pi \triangleq (R_0, Z_0, R_1, Z_1 \dots)$. This policy consists of two distinct sequences: (i) a sequence of routing decisions $(R_0, R_1, R_2, \dots)$, where $R_i$ specifies the route selected for transmitting the $(i+1)$-th packet, and (ii) a sequence of finite waiting times $(Z_0, Z_1, Z_2, \dots)$, where $Z_i < \infty$ determines the $(i+1)$-th sampling (or submission) time as $S_{i+1} = D_i + Z_i$. 

Let $\Pi$ denote the set of all causal and stationary policies $\pi$. We consider the following problem:

\begin{problem}[Average Age Penalty Minimization with Energy Constraint]\label{p1}
\begin{equation}
    \begin{aligned}
\lambda^\star
= &\min_{\pi \in \Pi}\;
 \limsup_{T \to \infty}\;
\frac{1}{T}\mathbb{E}\!\left[\int_{0}^{T} f(\Delta(t))\, dt \right]
\label{eq:obj-cts}
\\
\text{s.t.}\quad
& \limsup_{T \to \infty}\;
\frac{1}{T}\mathbb{E}\!\left[
\left(
\int_{0}^{T} E(t)\,dt
+ C_s\,N_s(T)
\right)
\right]\le E_{\max}
\end{aligned}
\end{equation}
where $N_s(T)$ is the number of sampling actions up to time $T$, and $\lambda^\star$ is the optimal long-term average AoI cost. This problem aims at minimizing the long-term average age penalty under a long-term average energy constraint.
\end{problem}

The AoI process $\Delta(t)$ is inherently a piecewise linear function as defined by equation~\eqref{aoidefinition}. Hence, it is natural to rewrite Problem~\ref{p1} as an average sum over the renewal intervals $[D_i, D_{i+1})$ corresponding to consecutive successful updates. 
The long-term time average in~\eqref{eq:obj-cts} can be expressed as
\begin{equation}
\label{eq8}
\begin{aligned}
        &\limsup_{T \to \infty} 
\mathbb{E}\!\left[ \frac{1}{T}\int_{0}^{T}\!f(\Delta(t))\,dt \right]
=\\& \limsup_{n \to \infty} 
\frac{ \mathbb{E}\!\left[ \frac{1}{n}\sum_{i=0}^{n-1} q(Y_{i}, Z_i, Y_{i+1}) \right] }
     { \mathbb{E}\!\left[ \frac{1}{n}\sum_{i=0}^{n-1} (Z_i + Y_{i+1}) \right]},
\end{aligned}
\end{equation}
where $q(Y_{i}, Z_i, Y_{i+1})$ represents the accumulated age penalty (the area under the curve $f(\Delta(t))$) during the $i$-th cycle. An illustrative example of this process for an exponential penalty function is shown in Fig.~\ref{fig_age_exp}, where the shaded region corresponds to the accumulated cost $q(\cdot)$.

In addition, the energy process $E(t)$ is also a piecewise constant function. Similarly, the long-term average energy consumed up to the $n$-th update can be written as: 
\begin{equation}\label{eq9}
\begin{aligned}
&\limsup_{T \to \infty}\;
\frac{1}{T}\mathbb{E}\!\left[
\left(
\int_{0}^{T} E(t)\,dt
+ C_s\,N_s(T)
\right)
\right] \\
&=\limsup_{n \to \infty} \frac{\mathbb{E}\!\left[ \frac{1}{n}\sum_{i=0}^{n-1} E_{i+1} \right]}{ \mathbb{E}\!\left[ \frac{1}{n}\sum_{i=0}^{n-1} (Z_i + Y_{i+1}) \right]},
\end{aligned}
\end{equation}
where $E_{i+1}$ is given in \eqref{eq.5}.

With \eqref{eq8} and \eqref{eq9}, Problem~\ref{p1} is written as:
\begin{problem}[Reformulated Non-linear Age Minimization]\label{eq1}
\begin{subequations}\label{eq1-}
        \begin{align}
\lambda^\star \triangleq \min_{\pi}\;
& \limsup_{n\to\infty}\;
\frac{ \mathbb{E}\!\left[\frac{1}{n}\sum_{i=0}^{n-1} q(Y_{i}, Z_i, Y_{i+1}) \right] }
     { \mathbb{E}\!\left[ \frac{1}{n}\sum_{i=0}^{n-1} (Z_i + Y_{i+1}) \right] }\label{obj}
\\\label{constraint}
\text{s.t.}\quad
& \limsup_{n\to\infty}\;
\frac{ \mathbb{E}\!\left[ \frac{1}{n}\sum_{i=0}^{n-1} E_{i+1} \right] }
     { \mathbb{E}\!\left[\frac{1}{n} \sum_{i=0}^{n-1} (Z_i + Y_{i+1}) \right] }
\;\le\; E_{\max}
\end{align}
\end{subequations}
\end{problem}

Problem \ref{eq1} is an infinite-horizon Constrained Semi-Markov Decision Process (CSMDP) with a hybrid state space and a state-dependent hybrid action space, posing significant challenges due to its fractional objective and the presence of a long-term energy constraint. To tackle this, we adopt a \textit{divide-then-concur} strategy, consisting of three key steps: $i$) \textit{Fractional constraint reformulation}: We first transform the original fractional constraint into an equivalent linear form to simplify the constraint structure; $ii$) \textit{Dinkelbach transformation}: Next, we apply the Dinkelbach approach \cite{dinkelbach1967nonlinear} to convert the fractional objective into a sequence of parameterized CMDP problems; $iii$) \textit{Lagrangian techniques}: Finally, we employ the Lagrangian method to handle the constraint in the Constrained Markov Decision Process (CMDP) formulation, effectively reducing the problem to a standard MDP.

\subsection{Fractional programming reformulation}

The constraint in Problem~\ref{eq1} involves a fractional form, which presents significant analytical and computational challenges. In particular, the presence of a long-run average ratio of expected costs makes the problem non-convex and difficult to tackle using standard dynamic programming. To address this issue, we first transform the fractional constraint into an equivalent linear form. The key observation is that the denominator in the fractional constraint (given in~\eqref{constraint}) is strictly positive under any admissible policy. This allows us to multiply both sides of the inequality by the denominator without changing the constraint’s validity. As a result, the fractional constraint in~\eqref{eq1-} can be equivalently rewritten as:
\begin{equation}
\begin{aligned}\label{eq:ineq-form}
\limsup_{n\to\infty}
\Big(
\mathbb{E}\left[ \frac{1}{n}\textstyle\sum_{i=0}^{n-1}(E_{i+1}-E_{\max}Z_i-E_{\max}Y_{i+1})  \right]
\le 0.
\end{aligned}
\end{equation}

Meanwhile, the objective in Problem~\ref{eq1} is a nonlinear fractional function of the policy~$\pi$, involving the ratio of two expected cumulative costs over an infinite horizon. Such fractional objectives are inherently non-convex and difficult to optimize directly, especially in the presence of additional constraints. To overcome this difficulty, we adopt \textit{Dinkelbach’s method}, a classic approach for solving nonlinear fractional programming problems. Specifically, for a given scalar parameter $\lambda \ge 0$, we define the following auxiliary problem:

\begin{problem}[Dinkelbach’s Reformulation]\label{p2}
\begin{subequations}\label{eq2}
\begin{align}
&h(\lambda) \triangleq \min_{\pi \in \Pi}\bigg\{\limsup_{n \to \infty}\notag\\
& 
\frac{1}{n} 
\sum_{i=0}^{n-1} 
\mathbb{E}\!\Big[
    q(Y_{i}, Z_,i, Y_{i+1})
    - \lambda (Z_i + Y_{i+1})
\Big]\bigg\}\\
&\text{s.t.}\limsup_{n\to\infty}
\Big(
\mathbb{E}\left[ \frac{1}{n}\textstyle\sum_{i=0}^{n-1}(E_{i+1}-E_{\max}(Z_i+Y_{i+1}))  \right]
\le 0
\end{align}
\end{subequations}
\end{problem}
The function $h(\lambda)$ measures the difference between the expected accumulated AoI cost and the weighted accumulated epoch duration. This problem is formulated as a CMDP, which we next transform into a standard MDP by utilizing Lagrangian techniques.

\subsection{Lagrangian Techniques}

Define the \textit{Lagrangian function} as:
\begin{equation}\label{eq:h_lambda_c_def}
\begin{aligned}
&h(\lambda,c) \triangleq 
\min_{\pi \in \Pi}\bigg\{ 
\limsup_{n \to \infty} 
\frac{1}{n} \sum_{i=0}^{n-1} 
\mathbb{E}\!\Big[
q(Y_{i}, Z_i, Y_{i+1})
\\&- \lambda (Z_i + Y_{i+1})
- c\big(
E_{\max}(Z_i + Y_{i+1})
- E_{i+1}
\big)
\Big]\bigg\},
\end{aligned}
\end{equation}
where the nonnegative Lagrange multiplier $c \ge 0$ penalizes violations of the long-term average energy constraint.

 The following lemma establishes the relationship between $h(\lambda,c)$ and the conditionally optimal long-term average AoI $\lambda_c^\star$ in the energy-constrained case.

\begin{Lemma}\label{lem1}
For any fixed $c \ge 0$, the following assertions hold:
\begin{enumerate}
\item $\lambda_c^\star \gtreqless \lambda$ if and only if $h(\lambda,c) \gtreqless 0$.
\item If $h(\lambda,c) = 0$, the solutions to the constrained Problem~\ref{eq1} and Problem~\ref{p2} coincide for the corresponding multiplier $c$.
\end{enumerate}
\end{Lemma}
\begin{proof}
The proof closely follows the derivation in \cite[Lemma 2]{8812616}, adapted here for the constrained Lagrangian formulation. See Appendix~\ref{appa}.
\end{proof}

Hence, for a fixed $c$, solving $h(\lambda,c)=0$ yields the conditionally optimal long-term average AoI $\lambda_c^\star$.  
Let \(E(\lambda_c^\star,c)\) denote the corresponding average energy consumption of the conditionally optimal policy which we denote by $\pi_c^\star$. Then, the average energy consumption is:
\begin{equation}\label{energy}
    E(\lambda_c^\star,c)
\triangleq\limsup_{n\to\infty}
\frac{\displaystyle
\mathbb{E}\!\left[
\frac{1}{n}\sum_{i=0}^{n-1}
\big(
G_{r_{i,c}^\star}\,Y_{i+1} + C_s
\big)
\right]}
{\displaystyle
\mathbb{E}\!\left[
\frac{1}{n}\sum_{i=0}^{n-1}
(
z_{i,c}^\star + Y_{i+1}
)
\right]},
\end{equation}
where $r_{i,c}^\star$ and $z_{i,c}^\star$ are $i$-th epoch routing and waiting decisions made by using $\pi_c^\star$.
Since $E(\lambda_c^\star,c)$ is monotonically decreasing in $c$ by complementary slackness~\cite{altman2021constrained}, the optimal Lagrange multiplier is obtained as
$c^\star = \inf\{\,c>0 : E(\lambda_c^\star,c) \le E_{\max}\,\}.$
At $(\lambda^\star,c^\star)$, both the average AoI and energy constraint are simultaneously optimal, satisfying $h(\lambda^\star,c^\star)=0$.

According to Lemma~\ref{lem1}, the solution to the energy-constrained problem in equation~\eqref{eq1-} can be obtained by identifying the value of $\lambda$ for which $h(\lambda,c)=0$ for a given multiplier $c$, and then optimizing over $c$. 
The root of the function $h(\lambda,c)$ thus corresponds to the optimal long-term average age $\lambda^\star$ under the average energy constraint, and the pair $(\lambda^\star,c^\star)$ jointly characterizes the optimal policy $\pi^\star$.

\subsection{Average-Cost MDP for a Given $\lambda$ and $c$}\label{sectionMDP}

In the third step, we show that Problem~\ref{p2} can be formulated as an average-cost Markov Decision Process (MDP) for a fixed value of $\lambda$, described by the quadruple
$\mathscr{M}(\lambda) \triangleq (\mathcal{S}, \mathcal{A}, \mathcal{P}, \mathcal{C}).$
Each component of this MDP is defined as follows.

\begin{itemize}
    \item \ul{\textit{State Space $\mathcal{S} = [0,\infty) \times \{0,1\}^N$}}:  
    At each decision epoch $i$, the system occupies a state 
    $(Y_{i}, \mathbf{L}_i) \in \mathcal{S}$, where the continuous component $Y_{i} = y < \infty$ denotes the delay observed in the previous transmission, and $\mathbf{L}_i = (l_{1,i}, l_{2,i}, \ldots, l_{N,i}) \in \{0,1\}^N$ encodes the availability of the $N$ routes at epoch $i$. Specifically,
    \begin{equation}
    l_{k,i} =
    \begin{cases}
        0, & \text{if route } k \text{ is available at time } i, \\
        1, & \text{otherwise}.
    \end{cases}
    \end{equation}

    \item \ul{\textit{Action Space $\mathcal{A}(\mathbf{L}_i)$}}:  
    Given the current availability vector $\mathbf{L}_i$, the decision maker selects an action $(R_i, Z_i)$, where
    \begin{itemize}
        \item $R_i\in\mathcal{R}(\mathbf{L}_i)$ denotes the chosen route for transmission, with $\mathcal{R}(\mathbf{L}_i)$ denoting the set of the available routes, defined as $\mathcal{R}(\mathbf{L}_i)=\{r|l_{k,i}=0\}$.
        \item $Z_i \in\mathbb{R}^+$ represents the waiting time before generating the next update.
    \end{itemize}
    The set of admissible actions depends on $\mathbf{L}_i$: $\mathcal{A}(\mathbf{L}_i) = \mathcal{R}_i\times \mathbb{R}^+$.
The global action space is defined as the union of all state-dependent sets: $\mathcal{A} \triangleq \bigcup_{\mathbf{L} \in \{0,1\}^N} \mathcal{A}(\mathbf{L})$.

\item \ul{\textit{State Transition Probability
$\mathcal{P}:\mathcal{S}\times\mathcal{A}\times \mathscr{B}(\mathcal{S}) \to [0,1]$}}:
Let $\mathscr{B}(\mathcal{S})$ denote the Borel $\sigma$-algebra generated by the measurable subsets of $\mathcal{S} = [0,\infty) \times \{0,1\}^N$.
For any current state $(y, \mathbf{l}) \in \mathcal{S}$, admissible action $(r, z) \in \mathcal{A}(\mathbf{l})$, and measurable set $C \in \mathscr{B}(\mathcal{S})$, the transition kernel is defined as
$
\mathcal{P}\big(C \mid y, \mathbf{l}, r, z \big)
= \mathbb{P}\big( (Y_{i+1}, \mathbf{L}_{i+1}) \in C
\mid Y_{i} = y, \mathbf{L}_i = \mathbf{l}, R_i = r, Z_i = z \big).
$ We assume that, conditioned on the current state and action,
the next delay $Y_{i+1}$ and the next availability vector $\mathbf{L}_{i+1}$ are \emph{conditionally independent}.
That is,
$
\mathbb{P}\big(Y_{i+1}, \mathbf{L}_{i+1} \mid Y_{i} = y, \mathbf{L}_i = \mathbf{l}, R_i = r, Z_i = z \big)
= \mathbb{P}(Y_{i+1} \mid R_i = r)  \mathbb{P}(\mathbf{L}_{i+1}),
$
meaning that the delay distribution depends only on the chosen route $r$, while the availability process evolves independently of the previous state or action. Hence, for any measurable set $C \subseteq \mathcal{S}$,  $\mathcal{P}\big(C \mid y, \mathbf{l}, r, z \big)
= \sum_{\mathbf{l}' \in \{0,1\}^N}
\mathbb{P}(\mathbf{L}_{i+1} = \mathbf{l}')
\int_{C_Y(\mathbf{l}')} Q_r(y') dy'$, where
$
C_Y(\mathbf{l}') := { y' \ge 0 : (y', \mathbf{l}') \in C }.
$
The availability process ${\mathbf{L}_i}$ is independent across epochs and identically distributed with stationary distribution $\mathbb{P}(\mathbf{L}_{i+1} = \mathbf{l}') =
\prod_{k=1}^N p_k^{1 - l_k'} (1 - p_k)^{l_k'}$.

    \item \ul{\textit{Cost Function $\mathcal{C}:\mathcal{S}\times\mathcal{A} \to \mathbb{R}$}}:  
    Let $F(t) \triangleq \int_{0}^{t} f(\tau) d\tau$ be the primitive of the penalty function $f(\cdot)$. The one-step cost incurred when the system is in state $(y, \mathbf{l})$ and action $(r, z)$ is taken is denoted by $g(y, \mathbf{l}, z, r; \lambda,c)$, defined as
    \begin{equation} \label{eq3}
        \begin{aligned}
             &g(y, \mathbf{l}, z, r; \lambda,c) \\
    &\triangleq \mathbb{E}_{Y\sim Q_r} \!\left[ q(y, z, Y) \right]+cC_s - (\lambda+cE_{\max}) z  \\&- \left(\lambda + cE_{\max} - cG_r\right)\, \mathbb{E}_{Y\sim Q_r}[Y] \\
    &= \mathbb{E}_{Y\sim Q_r} \!\left[ F(y + Y + z) - F(y) \right] + cC_s - (\lambda+cE_{\max}) z \\& - \left(\lambda + cE_{\max} - cG_r\right)\, \mu_r.
        \end{aligned}
    \end{equation}
    Here, the term $q(y, z, Y)$ represents the accumulated penalty area during the updating interval. 
\end{itemize}

Using Lemma~\ref{lem1} and the constructed MDP $\mathscr{M}(\lambda)$ for a fixed $\lambda$, we can design a \textit{nested} three-layer optimization algorithm (e.g., \cite{bedewy2021optimal, 10806969}) to solve the CMDP in Problem~\eqref{p1}. The details of this numerical solution are presented in Section~\ref{numericalsolutions}.

\section{Main Results}
To facilitate the numerical solution of Problem~\ref{p1}, we first derive several fundamental structural properties of the jointly optimal sampling and routing policies.

\subsection{Structural Results of Optimal Policies}
The introduction of general non-linear penalty functions $f(\cdot)$ generalizes the optimal policy structure. While the optimal waiting policy maintains a target-threshold structure for any monotonic non-decreasing $f(\cdot)$, the strict monotonicity of the routing policy is guaranteed only under certain conditions.

\begin{theorem} \label{theorem_main}
For an $N$-route problem where the delay distribution of each route has infinite support, the jointly optimal sampling and routing policies exhibit the following structure:

\begin{enumerate}
    \item \ul{Optimal Routing (Conditional):} 
    Let $\Psi_r(y) \triangleq \mathbb{E}_{Y \sim Q_r}[f(y + Y)]$ denote the expected immediate penalty on route $r$ given a current system delay $y$. If the penalty function $f(\cdot)$ and the delay distributions satisfy the \textit{Expected Penalty Ordering} property, defined as:
    \begin{equation} \label{eq:ordering_condition}
        \mu_j \ge \mu_k \implies \Psi_j(y) \ge \Psi_k(y) \quad \forall y \ge 0,
    \end{equation}
    then the optimal routing action $R_i^\star$ is a monotonic step function of the observed delay $Y_{i}$. Specifically, the routing decision is determined by $K$ positive thresholds such that the policy switches to routes with statistically "better" (lower mean) delay profiles as the previous delay $Y_{i}$ increases:
    \begin{equation} \label{eqrouting}
    R_i^{\star}(\mathbf{l}_i)=\sum_{k=1}^{K+1} \big ( a_k(\mathbf{l}_i)-a_{k-1}(\mathbf{l}_i) \big ) u\big(Y_{i}-\tau_{k-1}(\mathbf{l}_i)\big),
    \end{equation}
    where $a_k(\mathbf{l}_i)$ are indices of routes sorted by preference, and $u(t)$ is the unit step function. 
    
    \textit{Examples:} This sufficient condition is satisfied by:
    \begin{itemize}
        \item The \textbf{Linear} penalty $f(\Delta)=\Delta$ (since $\mu_j > \mu_k \implies y+\mu_j > y+\mu_k$).
        \item The \textbf{Exponential} penalty $f(\Delta)=e^{\alpha \Delta}-1$, provided the delay distributions satisfy the monotonicity of the moment generating function (i.e., $\mu_j > \mu_k \implies M_j(\alpha) > M_k(\alpha)$).
    \end{itemize}
        \begin{remark}
            For general convex functions (e.g., $f(\Delta)=\Delta^2$), this monotonicity is not guaranteed due to the coupling of current age with higher-order delay moments.
        \end{remark} 
    
    \item \ul{Optimal Sampling (General):} For any monotonic non-decreasing penalty function $f(\cdot)$, the optimal waiting time at the $i$-th epoch, $Z_i^\star$, follows a target-threshold structure. Specifically, for a chosen route $R_i$, there exists an optimal target age $\beta^\star(R_i)$ such that:
    \begin{equation} \label{optimal_waiting_eq}
        Z_i^{\star}(\mathbf{l}_i) = \left( \beta^\star(R_i) - Y_{i} \right)^+,
    \end{equation}
    where $(\cdot)^+\triangleq\max\{0,\cdot\}$. The target age threshold $\beta^\star(R_i)$ is the unique solution to the following stopping rule:
    \begin{equation} \label{eq:beta_star}
    \begin{aligned}
        &\beta^\star(R_i) =\\& \inf \left\{ t \ge 0 : \mathbb{E}_{Y}\left[ f(t + Y_{i+1}) \right] \ge \lambda^\star + c^\star E_{\max} \right\},
    \end{aligned}
    \end{equation}
    where $Y_{i+1}$ is the random delay of the chosen route, $\lambda^\star$ is the optimal average AoI penalty, and $c^\star$ is the Lagrange multiplier associated with the energy constraint.
    
    This implies the system waits until the \textit{expected penalty at the moment of delivery} reaches the critical system cost $\lambda^\star + c^\star E_{\max}$. In the special case where $f(\Delta)=\Delta$ (linear), \eqref{eq:beta_star} simplifies to $\beta^\star(R_i) = \lambda^\star + c^\star E_{\max} - \mu_{R_i}$, recovering the standard result.
\end{enumerate}
\end{theorem}
\begin{proofsketch}
    With the MDP $\mathscr{M}(\lambda)$ defined by the cost function in \eqref{eq3}, we can establish the \textit{Average-Cost Optimality Equation} (ACOE) \cite[Eq. 4.1]{howard1960dynamic}:
    \begin{equation}\label{eq4}
    \begin{aligned}
        V^*(y,\mathbf{l}; \lambda,c) + &h(\lambda,c) = \min_{(z, r)\in \mathcal{A}(\mathbf{l})} \bigg\{ g(y, \mathbf{l}, z, r; \lambda,c) \\&+ \mathbb{E}_{Y'\sim Q_r,\mathbf{l}'\sim\boldsymbol{p}} \big[ V^*(Y',\mathbf{l}'; \lambda,c) \big] \bigg\},
    \end{aligned}
\end{equation}
where $V^*(y,\mathbf{l}; \lambda,c)$ is the relative value function, and $h(\lambda,c)$ is the optimal value of the reformulated MDP in Problem \ref{p2}. Given any $\lambda$, $c$ and route $r\in\mathcal{N}$, we first prove that the optimal waiting time that solves the right hand-side of \eqref{eq4} follows a \textit{target-threshold} structure, given by:
\begin{equation}\label{zyrlambda}
    z^\star(y,\mathbf{l};r,\lambda,c)=(\beta^\star(r, \lambda, c)-y)^+,
\end{equation}
where $\beta^\star(r, \lambda, c)$ is the route-specific target age derived from the stopping rule given in \eqref{eq:beta_star}.

As $h(\lambda_c^\star,c)=0$ for the conditionally optimal $\lambda_c^\star$ for any $c$, applying $\lambda=\lambda_c^\star$ in \eqref{eq4} and \eqref{zyrlambda} yields:
    \begin{equation}\label{acoe2}
    \begin{aligned}
         V^*(y,\mathbf{l}; \lambda_c^\star,c) = \min_{r \in \mathcal{R}(\mathbf{l})} \Big\{ &g(y, \mathbf{l}, z^\star(y,\mathbf{l};r,\lambda_c^\star,c), r; \lambda_c^\star,c) \\ &+\mathbb{E}_{Y'\sim Q_r,\mathbf{l}'\sim\boldsymbol{p}} \big[ V^*(Y',\mathbf{l}'; \lambda_c^{\star},c) \big]\Big\}. 
    \end{aligned}
    \end{equation}
    For short-hand notations, we define the action-value  as:
    \begin{equation}\label{actionvalue}
    \begin{aligned}
        Q(y,\mathbf{l},r)\triangleq &g\big(y, \mathbf{l},z^\star(y;r,\lambda_c^\star,c), r; \lambda_c^\star,c\big) \\ &+ \mathbb{E}_{Y'\sim Q_r,\mathbf{l}'\sim\boldsymbol{p}} \big[ V^*(Y',\mathbf{l}'; \lambda_c^{\star},c) \big],
    \end{aligned}
    \end{equation}
    and the optimal routing policy $r^\star(y,\mathbf{l})$ turns to
    \begin{equation}
        r^\star(y,\mathbf{l})=\operatorname*{arg\,min}_{r \in \mathcal{R}(\mathbf{l})}\{Q(y,\mathbf{l},r)\}.
    \end{equation}
    Then, we analyze a series of properties of the function $Q(y,\mathbf{l},r)$ and prove that, under the Expected Penalty Ordering property, $r^\star(y,\mathbf{l})$ is a non-decreasing step function. The full derivation of this result is provided in Section \ref{SecPfMain}.
\end{proofsketch}

The following Lemma \ref{lem_thrbound} demonstrates an important relationship between the sampling target threshold $\beta_k^\star(\mathbf{l}_i)$ (associated with the $k$-th preferred route $a_k$) and the routing switching threshold $\tau_k(\mathbf{l}_i)$.

\begin{figure}[t] 
    \centering
\includegraphics[width=0.7\linewidth]{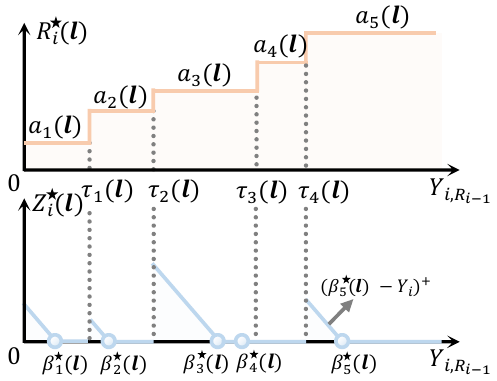} 
    \caption{Visualization of the jointly optimal policies.}
    \label{fig_pol}
\end{figure}

\begin{Lemma} \label{lem_thrbound}
    For penalty functions yielding a monotonic routing structure, the following assertion holds true:
    \begin{equation}
        \beta_k^\star(\mathbf{l}_i) < \tau_k(\mathbf{l}_i), \quad k \le K.
    \end{equation}
    This implies that the optimal policy always terminates the waiting phase before the age accumulates to the level where a route switch would become optimal.
\end{Lemma}
\begin{proof}
    See Appendix~\ref{proof:lemma1}.
\end{proof}
Consequently, for such penalty functions, for any interval $Y_{i}\in \big[\tau_{k-1}(\mathbf{l}_i),\tau_{k}(\mathbf{l}_i)\big)$ associated with a constant optimal routing option $a_k(\mathbf{l}_i)$, there exists a corresponding sub-interval $\big[\beta^\star_k(\mathbf{l}_i),\tau_k(\mathbf{l}_i)\big)$ in which a zero-waiting policy, defined by $Z_i^\star(\mathbf{l}_i)=\big(\beta_k^\star(\mathbf{l}_i)-Y_{i}\big)^+=0$, is optimal. Figure~\ref{fig_pol} provides an example sketch that illustrates the structure of the jointly optimal sampling and routing policy under a given availability state $\mathbf{l}$. 
The threshold-based structure derived in this subsection enables highly efficient deployment in complex networks. Terminals can maximize information \textit{freshness} simply by storing and applying the derived thresholds.
In Section~\ref{numericalsolutions}, we present a series of algorithms to compute these thresholds efficiently.

\subsection{Bounds on the Optimal Average AoI Penalty}
In this subsection, we establish the upper and lower bounds on the optimal average age penalty $\lambda^\star$. These bounds will later serve as initialization points for the bisection search described in Section \ref{numericalsolutions}.

\begin{Lemma}\label{lembounds}
$\lambda^\star$ is bounded by:
\begin{equation}
\begin{aligned}
    &f(0) \le \lambda^\star \\&\le \min_{k\in \mathcal{R}_{\infty}} \left( \frac{\mathbb{E}_{Y, Y' \sim Q_k} [ F(Y' + w_k + Y) - F(Y') ]}{\mu_k + w_k} \right) = \lambda^u,
\end{aligned}
\end{equation}
where $F(t) \triangleq \int_0^t f(\tau) d\tau$, $Y$ and $Y'$ are independent random variables representing consecutive delays on route $k$, and $w_k$ is the minimum constant waiting time required to satisfy the energy constraint on route $k$: $w_k = \left( \frac{C_s + G_k\mu_k}{E_{\max}} - \mu_k\right)^+$.
\end{Lemma}
\begin{proof}
See Appendix~\ref{ap3.2.4}.
\end{proof}

\noindent \textit{Remark:} The upper bound corresponds to the performance of a naive stationary policy that exclusively uses route $k$ with a constant waiting time $w_k$. In the special case of linear AoI ($f(t)=t$), the numerator simplifies to $\frac{\mathbb{E}[(Y'+w_k+Y)^2] - \mathbb{E}[Y'^2]}{2}$, recovering the explicit bound $\frac{3\mu_k + w_k}{2}+\frac{\sigma_k^2}{2(\mu_k + w_k)}$.

\section{Numerical Solutions}\label{numericalsolutions}
In this section, we develop numerical algorithms to solve the energy-constrained average age minimization problem in~\ref{p1} and determine the thresholds introduced in Theorem~\ref{theorem_main}. By leveraging the Lagrangian relaxation and Dinkelbach’s method introduced in Section~\ref{section2.4}, the problem can be reformulated as a two-layer nested structure. 

In the \textit{inner layer}, for a fixed pair $(\lambda,c)$, we approximate the auxiliary function $h(\lambda,c)$ defined in~\eqref{eq2} through the Average Cost Optimality Equation (ACOE)~\eqref{eq4}. In the \textit{outer layer}, we update $(\lambda,c)$ iteratively until the Dinkelbach equilibrium $h(\lambda^\star,c^\star)=0$ and the energy constraint are both simultaneously satisfied. 

\subsection{Challenges in Approximating $h(\lambda,c)$}

\subsubsection{Challenge 1: Hybrid Action Space} 
The first challenge in computing $h(\lambda,c)$ arises from the \textit{hybrid} nature of the action space, where sampling actions $z\in\mathbb{R}^+$ are continuous, while routing actions $r\in\mathcal{R}$ are discrete and depend on the current route availability. By leveraging~\eqref{zyrlambda}, which gives the optimal sampling policy $z^\star(y,\mathbf{l},r;\lambda,c)=(\beta^\star(r, \lambda, c)-y)^+,$
we substitute $z^\star(\cdot)$ into the ACOE~\eqref{eq4}. This transformation yields a simplified SMDP with a countable routing action space:
\begin{equation}\label{eq12v2_energy}
\begin{aligned}
    V^*(y,\mathbf{l}; \lambda,c) + h(\lambda,c)
    &= \min_{r\in\mathcal{R}(\mathbf{l})} 
    \big\{
    g(y,\mathbf{l}, z^\star(y;\lambda,c), r; \lambda,c) \\
      &+ \mathbb{E}_{Y'\sim Q_r,\mathbf{l}'\sim\boldsymbol{p}}
    [\,V^*(Y',\mathbf{l}'; \lambda,c)\,]
    \big\}.
\end{aligned}
\end{equation}
This effectively decouples the continuous and discrete components of the hybrid action space, reducing the dimensionality of the optimization.

\subsubsection{Challenge 2: Uncountable State Space} 
The second challenge stems from the uncountable nature of the continuous state variable $y\in\mathbb{R}^+$. Evaluating~\eqref{eq12v2_energy} over the entire state space is computationally intractable. Traditional discretization methods approximate $\mathcal{S}$ by a finite grid of $M$ points $\{y_1,\ldots,y_M\}$, introducing a quantization error $\epsilon_M$ that vanishes asymptotically with $M$. The overall complexity of the Relative Value Iteration (RVI) approach scales as $\mathcal{O}(|\mathcal{N}|M^2)$.

\subsection{Proposed \textsc{ReaVI} Algorithm with Energy Constraint}

To overcome the trade-off between accuracy and computational cost, we extend the \textbf{R}elative \textbf{E}xpected \textbf{A}ction \textbf{V}alue \textbf{I}teration (\textsc{ReaVI}) algorithm to handle the Lagrangian form of the constrained problem. Define the relative value function
\begin{equation}\label{eq10_energy}
W^*(y,\mathbf{l};\lambda,c)
\triangleq V^*(y,\mathbf{l};\lambda,c) - V^*(0,\mathbf{l};\lambda,c),
\end{equation}
where $W^*(0,\mathbf{l};\lambda,c)=0$.
Substituting~\eqref{eq10_energy} into~\eqref{eq12v2_energy} gives
\begin{equation}\label{eq12_energy}
\begin{aligned}
    &W^*(y,\mathbf{l}; \lambda,c) + h(\lambda,c)
=\\& \min_{r\in\mathcal{R}(\mathbf{l})} 
\Big\{
g(y,\mathbf{l}, z^\star(y;r,\lambda,c), r; \lambda,c)
+ \\&\mathbb{E}_{Y'\sim Q_r,\mathbf{l}'\sim\boldsymbol{p}}
[\,W^*(Y',\mathbf{l}'; \lambda,c)\,]
\Big\}.
\end{aligned}
\end{equation}
Define the \textbf{R}elative \textbf{E}xpected \textbf{A}ction \textbf{V}alue (\textsc{ReaV}) function as follows:
\begin{equation}\label{eq14v2_energy}
G(r;\lambda,c)
\triangleq
\mathbb{E}_{Y\sim Q_r,\mathbf{l}\sim\boldsymbol{p}}
\big[ W^*(Y,\mathbf{l}; \lambda,c) \big],
\quad r\in\mathcal{N}.
\end{equation}
This converts the uncountable state space $\mathbb{R}^+$ into a fixed-point problem over the finite route set $\mathcal{N}$. Taking the expectation over the random variables $(Y,\mathbf{l})$ yields the \textit{Relative Expected Action Value Optimality Equation (REAVOE)}:
\begin{equation}\label{reavoe_energy}
\begin{aligned}
    &G(q;\lambda,c)+h(\lambda,c)=\mathbb{E}_{Y\sim Q_q,\mathbf{l}\sim\mathbf{p}}\\&\left[\min_{r\in\mathcal{R}(\mathbf{l})}
\big\{g(Y,\mathbf{l}, z^\star(Y;r,\lambda,c), r; \lambda,c)+G(r;\lambda,c)\big\}\right],
\end{aligned}
\end{equation}
with
\begin{equation}\label{h_lambda_c}
\begin{aligned}
    &h(\lambda,c)
\\&=
\mathbb{E}_{\mathbf{l}\sim\boldsymbol{p}}
\left[
\min_{r\in\mathcal{R}(\mathbf{l})}
\big\{
g(0,\mathbf{l}, z^\star(0;r,\lambda,c), r; \lambda,c)
+ G(r;\lambda,c)
\big\}
\right].
\end{aligned}
\end{equation}

\textbf{Algorithmic Solution:}
The resulting \textsc{ReaVI} algorithm iteratively updates $\{G(r;\lambda,c)\}_{r\in\mathcal{N}}$ and $h(\lambda,c)$ until convergence. In the middle layer, the Dinkelbach update for $\lambda$ is done via bisection search for a fixed $c$. In the outer layer, a bisection search for $c^\star$ is done to enforce the energy constraint. Given $(\lambda,c)$ and $\{G(r;\lambda,c)\}$, the selection of the one-step route is:
\begin{equation}\label{policy_argmin}
\begin{aligned}
    r^\star(y,\mathbf{l};\lambda,c)\in
\operatorname*{arg\,min}_{r\in\mathcal{R}(\mathbf{l})}
\big\{
&g(y,\mathbf{l}, z^\star(y;r,\lambda,c), r; \lambda,c)
\\&+ G(r;\lambda,c)
\big\}.
\end{aligned}
\end{equation}

\begin{algorithm}[t]
	\caption{\textit{Route Thresholds Algorithm}}
	\label{alg:thr_opt}
	\LinesNumbered
	\KwIn{$\lambda, \mathbf{l}, \{\mathcal{T}_{a,b}\}_{a \neq b},\{G(r;\lambda,c)\}_{r\in\mathcal{N}:\;l_r=0}$}
	\textbf{Initialize:} \\
	$i \leftarrow 1$; \\
    $\tau_{0}(\mathbf{l}) \leftarrow 0$\;
	$a_1(\mathbf{l}) = \arg\min_{r\in \mathcal{R}(\mathbf{l})} \{g(0,\mathbf{l},z^\star(0;r,\lambda,c),r;\lambda,c)+G(r;\lambda,c)\}$ \tcp*{Optimal route for $y=0$}
	$a_{\text{old}} \leftarrow a_1(\mathbf{l})$\;
	
	\While{\text{true}}{
		\tcp{Find the set of all future potential switching points}
		$\mathcal{P} \leftarrow \{ (t, r) : r \in \mathcal{R}(\mathbf{l}), r \neq a_{\text{old}}, t \in \mathcal{T}_{a_{\text{old}}, r}, \; t > \tau_{i-1}(\mathbf{l}) \}$\;
		
		\If{$\mathcal{P} = \emptyset$}{
			\textbf{Break} \tcp*{No more crossovers; $a_{\text{old}}$ is optimal for $y \in [\tau_{i-1}, \infty)$}
		}

		\tcp{Identify the very next threshold and the route to switch to}
		$\tau_i(\mathbf{l}) \leftarrow \min \{ t : (t, r) \in \mathcal{P} \}$\;
		$a_{i+1}(\mathbf{l}) \leftarrow \text{route } r \text{ corresponding to } \min \mathcal{P}$\;
		
		$a_{\text{old}} \leftarrow a_{i+1}(\mathbf{l})$\;
		$i \leftarrow i+1$\;
	}
	\KwOut{$\{\tau_j(\mathbf{l)}\}_{j=1}^{i-1}$ and $\{a_j(\mathbf{l)}\}_{j=1}^{i}$}
\end{algorithm}

\begin{algorithm}
\caption{\textit{Energy-Constrained ReaVI with Nested Bisection on $(\lambda,c)$}}
\label{alg:bisection_reavi}
\LinesNumbered
\KwIn{$E_{\max}$, bounds $c^-=0$, $c^+>0$, tolerances $\epsilon_\lambda,\epsilon_c,\epsilon_{\text{fp}}>0$}
\While{$c^+ - c^- > \epsilon_c$}{
  \tcp*[l]{Outer bisection on $c$}
  $c \leftarrow \dfrac{c^- + c^+}{2}$\;

  \tcp*[l]{Inner bisection on $\lambda$ (Dinkelbach root)}
  choose $\lambda^- = 0$ and $\lambda^+ = \lambda^u$\;

  \While{$\lambda^+ - \lambda^- > \epsilon_\lambda$}{
    $\lambda \leftarrow \dfrac{\lambda^-+\lambda^+}{2}$\;

    \tcp*[l]{REAVI fixed-point for given $(\lambda,c)$}
    initialize $G(r;\lambda,c)\leftarrow 0$ for all $r$\;
    $h \leftarrow
      \mathbb{E}_{\boldsymbol{p}}
      \Big[
      \min_{r\in\mathcal{R}(\mathbf{l})}
      \big\{ g\big(0,\mathbf{l}, z^\star(0;r,\lambda,c), r; \lambda,c\big)
      + G(r;\lambda,c) \big\}
      \Big]$\;

    \Repeat{$\bigl|h-h_{\text{old}}\bigr| < \epsilon_{\text{fp}}$}{
      $h_{\text{old}} \leftarrow h$\;

    \For{$q\in\mathcal{R}$}{
      \(
      \begin{aligned}
        &G(q) \leftarrow {} \\& -\,h_{\text{old}}
        + \mathbb{E}_{Y\sim Q_q,\mathbf{l}\sim\mathbf{p}}
          \Big[
            \min_{r\in\mathcal{R}(\mathbf{l})}
            \big\{
              g\big(Y,\mathbf{L},\\& z^\star(Y;r,\lambda,c), r; \lambda,c\big)
              + G(r;\lambda,c)
            \big\}
          \Big]
          ;
      \end{aligned}
      \)
    }

      $h \leftarrow \mathbb{E}_{\boldsymbol{p}}
      \Big[
      \min_{r\in\mathcal{R}(\mathbf{l})}
      \big\{ g\big(0,\mathbf{l}, z^\star(0;r,\lambda,c), r; \lambda,c\big)
      + G(r;\lambda,c) \big\}
      \Big]$\;
    }

    \tcp*[l]{Evaluate the Dinkelbach sign}
    set $h(\lambda,c)\leftarrow h$\;
    \lIf{$h(\lambda,c)>0$}{$\lambda^- \leftarrow \lambda$}
    \lElse{$\lambda^+ \leftarrow \lambda$}
  }

  Compute $\tau_{a,b}$'s and obtain $\mathcal{T}_{a,b}$'s\;
  Run Algorithm~\ref{alg:thr_opt} and find the policy $r^\star(y,\mathbf{l})$ in~\eqref{policy_argmin}\;
  compute $E(\lambda,c)$ by~\eqref{energy}\;
  \lIf{$E(\lambda,c)\ge E_{\max}$}{$c^- \leftarrow c$}
  \lElse{$c^+ \leftarrow c$}
}
Compute the average epoch lengths $\bar{T}^+$ and $\bar{T}^-$ of $\pi_{c^+}$ and $\pi_{c^-}$\;
Compute target time-fraction $\alpha \leftarrow 
\frac{E_{\max}- E(\lambda,c^-)}{ E(\lambda,c^+)- E(\lambda,c^-)}\in[0,1]$\;
Compute randomization probability:
$q \leftarrow \frac{\alpha \bar{T}^-}{\alpha \bar{T}^- + (1-\alpha)\bar{T}^+}$
Define $\pi^\star$ as choosing $\pi_{c^+}$ with probability $q$ and $\pi_{c^-}$ with probability $1-q$ at the start of each cycle\;

\KwOut{$\lambda^\star,c^\star,\pi^\star$, and $\{G(r;\lambda^\star,c^\star)\}_{r\in\mathcal{R}}$}
\end{algorithm}

    \section{Proof of Optimal Threshold Structures} \label{SecPfMain}
For short-hand notations, we define $Q(y,\mathbf{l},z,r;\lambda,c)$ as the \textit{state-action function} in the right-hand side of \eqref{acoe2}:
\begin{equation}\label{q1function}
    \begin{aligned}
        Q(y,\mathbf{l},z,r;\lambda,c)\triangleq &g(y,\mathbf{l},z,r; \lambda,c) \\&+ \mathbb{E}_{Y'\sim Q_r,\mathbf{l}'\sim \boldsymbol{p}}[V^*(Y',\mathbf{l}';\lambda,c)].
    \end{aligned}
\end{equation}
Given a specific route $r$, we determine the conditionally optimal waiting time $z^\star(y,\mathbf{l};r,\lambda,c)$ by analyzing the partial derivative of $Q$ with respect to $z$. Differentiating the objective function with respect to $z$ yields the expected marginal cost of waiting:
\begin{equation}\label{zd_nonlinear}
    \frac{\partial Q(y,\mathbf{l},z,r;\lambda,c)}{\partial z} = \mathbb{E}_{Y'\sim Q_r}[f(y+z+Y')] - (\lambda + cE_{\max}).
\end{equation}
Since $f(\cdot)$ is a monotonic non-decreasing function, the term $\mathbb{E}_{Y'}[f(y+z+Y')]$ is monotonically non-decreasing in $z$. We can therefore identify the optimal waiting time by comparing the expected penalty against the cost threshold $\lambda + cE_{\max}$:

\begin{itemize}
    \item \textbf{Case 1}: If $\mathbb{E}_{Y'}[f(y+Y')] \ge \lambda + cE_{\max}$, then for all $z \ge 0$, the derivative in \eqref{zd_nonlinear} is non-negative. This implies that the cost function is increasing with respect to $z$, and thus the optimal waiting time is $z^\star(y,\mathbf{l};r,\lambda,c) = 0$.
    
    \item \textbf{Case 2}: If $\mathbb{E}_{Y'}[f(y+Y')] < \lambda + cE_{\max}$, the derivative starts negative at $z=0$. Due to the monotonicity of $f(\cdot)$, there exists a unique target value, which we denote momentarily as $t^\star$, such that $\mathbb{E}_{Y'}[f(t^\star + Y')] = \lambda + cE_{\max}$. In this regime, $Q(y,\mathbf{l},z,r;\lambda,c)$ decreases until $y+z$ reaches $t^\star$, after which it increases. Consequently, the optimal waiting time is chosen such that the sampling occurs when the current age reaches this target, i.e., $z^\star = t^\star - y$.
\end{itemize}

We can unify these two cases by defining the route-specific target age $\beta^\star(r)$ using the following formulation:
\begin{equation} \label{beta_def}
    \beta^\star(r) = \inf \left\{ t \ge 0 : \mathbb{E}_{Y'\sim Q_r}[f(t + Y')] \ge \lambda + cE_{\max} \right\}.
\end{equation}
This definition covers both scenarios: in Case 2, $\beta^\star(r)$ is exactly the unique root $t^\star$; in Case 1, the condition holds for $t=0$ (or implies $\beta^\star(r) \le y$), resulting in a target that has already been exceeded. Thus, the optimal policy is described by the target-threshold structure:
\begin{equation} \label{eqwaiting}
    z^\star(y,\mathbf{l};r,\lambda,c) = (\beta^\star(r) - y)^+.
\end{equation}
				Substituting \eqref{eqwaiting} into \eqref{q1function} and setting $\lambda=\lambda^\star$ yields a compact form of $Q(y,\mathbf{l},r)$, whose definition has been given in \eqref{actionvalue}: 
\begin{equation}\label{eq99}
    \begin{aligned}
        Q(y,\mathbf{l},r) = & \; c C_s - (\lambda^\star + c E_{\max}) \left( (\beta^\star(r) - y)^+ + \mu_r \right) \\
        & + \mathbb{E}_{Y'\sim Q_r} \left[ F(y + (\beta^\star(r) - y)^+ + Y') - F(y) \right] \\
        & + \mathbb{E}_{Y'\sim Q_r,\mathbf{l}'\sim\boldsymbol{p}} \big[ V^*(Y',\mathbf{l}'; \lambda^\star, c) \big],
    \end{aligned}
\end{equation}
where $\mu_r = \mathbb{E}[Y']$ is the expected delay of route $r$, and the term $F(y + z^\star + Y') - F(y)$ captures the exact non-linear age penalty accumulated during the $i$-th interval.
With the notation $Q(y,\mathbf{l},r)$, the ACOE turns to:
\begin{equation}
    V^*(y,\mathbf{l}; \lambda^\star,c) = \min_{r\in \mathcal{R}(\mathbf{l})} \{ Q(y,\mathbf{l},r)  \}.
\end{equation}
Meanwhile, the optimal routing policy is given by:
\begin{equation}
    r^\star(y,\mathbf{l})=\operatorname*{arg\,min}_{r\in\mathcal{R}(\mathbf{l})}\{Q(y,\mathbf{l},r)\}.
\end{equation}

To analyze the threshold structure of $r^\star(y,\mathbf{l})$, the following lemmas discuss some important properties of $Q(y,\mathbf{l},r)$ and $V^*(y,\mathbf{l};\lambda^{\star},c)$. We begin with properties that hold for general monotonic non-decreasing penalty functions.

\begin{Lemma}\label{lemQsimplified}
The action-value function $Q(y,\mathbf{l},r)$ is independent of the state $\mathbf{l}$, i.e., $Q(y,\mathbf{l},r) = Q(y,r), \; \forall r \in \mathcal{R}$.
Hence, the optimality equation simplifies to:
\begin{equation}
    V^*(y,\mathbf{l}; \lambda^\star,c) = \min_{r\in \mathcal{R}(\mathbf{l})} \{ Q(y,r)  \}.
\end{equation}
\begin{proof}
    Even though $V^*(y,\mathbf{l}; \lambda^\star,c)$ depends on $\mathbf{l}$, once route $r$ has been selected, the term $\mathbb{E}_{Y'\sim Q_r,\mathbf{l}'\sim\boldsymbol{p}} \big[ V^*(Y',\mathbf{l}'; \lambda^\star,c) \big]$ in the right-hand side of \eqref{eq99} is independent of $\mathbf{l}$. 
\end{proof} 
\end{Lemma}

\begin{Lemma}\label{lem2}
For any monotonic non-decreasing penalty function $f(\cdot)$, the following assertions hold true: 
\begin{enumerate}
    \item $\forall r\in\mathcal{N}$, $Q(y,r)$ is monotonically increasing with $y$.
    \item For a given $\mathbf{l}$, $V^*(y,\mathbf{l}; \lambda^\star,c)$ is monotonically increasing with $y$.
\end{enumerate}
\end{Lemma}
\begin{proof}
    See Appendix~\ref{prooflemma5}.
\end{proof}

We now refine our analysis for the case of \textit{ordered expected penalties}, as defined in \eqref{eq:ordering_condition}. In this scenario, the ordering of the action-value derivatives is strictly determined by the route mean delays.

\begin{Lemma}\label{lem3}
Under the expected penalty ordering condition \eqref{eq:ordering_condition}, for any two routes $j, k \in \mathcal{N}$ such that $\mu_j > \mu_k$, we have
\begin{equation} \label{lem3eq}
    \frac{\partial Q(y,j)}{\partial y}\geq\frac{\partial Q(y,k)}{\partial y}, \quad \forall y\in\mathbb{R}^+. 
\end{equation}
\end{Lemma}
\begin{proof}
    See Appendix~\ref{prooflemma6}.
\end{proof}

With Lemma \ref{lem3} established, we can derive the threshold-based structure of the optimal routing policy for these specific penalty functions.

\begin{Lemma} \label{cor1}
Consider $N$ routes with their mean delays satisfying $\mu_1\ge\mu_2\cdots\ge\mu_N$. Under the expected penalty ordering condition \eqref{eq:ordering_condition}, if route $j$ is optimal at $y=y^\star$ for a given availability $\mathbf{l}$, then:
\begin{equation}
    \begin{aligned}
        r^\star(y,\mathbf{l})=\operatorname*{arg\,min}_{r\in\mathcal{R}(\mathbf{l})}\{Q(y,r)\}\ge j,\quad\text{if } y> y^\star,\\ 
        r^\star(y,\mathbf{l})=\operatorname*{arg\,min}_{r\in\mathcal{R}(\mathbf{l})}\{Q(y,r)\}\le j,\quad\text{if } y< y^\star.
    \end{aligned}
\end{equation}
\end{Lemma}
\begin{proof}
    See Appendix~\ref{lemma6:proof}.
\end{proof}

Lemma \ref{cor1} implies that under the Expected Penalty Ordering property, $r^\star(y,\mathbf{l})$ is a monotonically non-decreasing step function with respect to $y$. Consequently, the optimal sampling policy $z^\star(y) = (\beta^\star(r^\star(y,\mathbf{l})) - y)^+$ also follows a piecewise structure dictated by the changes in the optimal route.

Finally, we prove that there are at most $\binom{N}{2}$ unique routing thresholds for these specific cases. For any pair of routes $a, b \in \mathcal{N}$ with $a < b$, let $\mathcal{T}_{a,b}$ be the set of ages $y$ satisfying $Q(y,a) = Q(y,b)$. By Lemma \ref{lem3} and Lemma \ref{cor1}, if such a crossing exists, it is unique (i.e., $|\mathcal{T}_{a,b}| \le 1$). Thus, we simplify the notation by defining the scalar $\tau_{a,b}$ as the unique element of this set. Since there are at most $\binom{N}{2} = \frac{N(N-1)}{2}$ such pairs, the optimal routing policy is characterized by at most $\binom{N}{2}$ switching thresholds. Note that the existence of a threshold $\tau_{a,b}$ does not necessitate its presence in the optimal solution. 

For general monotonically non-decreasing penalty functions allowing at most $m$ crossover points between any pair of routes, the policy is characterized by at most $m\binom{N}{2}$ thresholds. We expect neither $m$ nor the overall number of thresholds to be large for practical penalty functions and delay distributions.

\section{Simulation Results}
	This section presents simulation results for practical scenarios to validate the analytical findings and evaluate the performance of our proposed algorithm. To provide a clear understanding of the system dynamics, we analyze the impact of non-linear age penalty functions, energy constraints, and stochastic route availability in isolation. This decoupled approach is adopted for the following reasons:

    First, mixing strict energy constraints with other system parameters can obscure the fundamental benefits of dynamic routing. As we observe in the subsequent analysis, in low-energy regimes, the optimal policy converges to a static single-route strategy to conserve power. By isolating the energy constraint, we can distinctly characterize the transition from energy-limited static behavior to diversity-driven dynamic routing, without confounding the results with non-linear penalty dynamics.

    Second, the analysis of stochastic route availability requires careful benchmarking. If the proposed policy is compared against a single-route baseline that is always available, the relative performance gain can become arbitrarily large, rendering the comparison trivial. By treating availability separately and proposing our own specific benchmarks for that case, we ensure fair and consistent comparisons against the standard always-available single-route policies established in prior literature for other constraints, allowing us to quantify the precise value added by the proposed algorithm under each unique practical constraint.

	\subsection{Comparing Benchmarks} \label{benchmark}
	In this subsection, we refer to our jointly optimal sampling and routing policy as the ``optimal policy'' and evaluate its performance against the following benchmark policies:
    
    \noindent $\bullet$ \textit{Minimum Average Delay Routing with AoI-Optimal Sampling (\textbf{MAD-Optimal})}: This policy always selects the route with the minimum average delay over the set of available routes at each instance. Given this selection, a modified version of the \textit{ReAVI} with the minimization over the routing options is implemented to find the AoI-optimal waiting strategy and minimize the long-term average AoI. This policy follows the AoI-optimal waiting strategy as outlined in \cite[Theorem 4]{sun2017update}, when $\boldsymbol{p} = \boldsymbol{1}$.

    \noindent $\bullet$ \textit{Minimum Average Delay Routing with Zero-Wait Sampling (\textbf{MAD-Zero Wait})}: This policy always selects the route with the minimum average delay over the set of available routes at each instance. It is combined with a zero-wait strategy, where a new packet is sampled and transmitted immediately upon the delivery of the previous packet. The long term average AoI achieved by this policy can be analytically calculated.
    \begin{Lemma} \label{lem-mad-zw-age}
    For a system with $N$ routes satisfying $\mu_1 \ge \mu_2 \ge \dots \ge \mu_N$, MAD-ZW policy under no energy constraint given by:
    \begin{equation}\label{pol-mad-zw-age}
        \pi^{\text{MAD-ZW}}(y,\mathbf{l})
        \triangleq \left(r = \max\{\,k \mid l_k = 0\,\},\; z = 0\right).
    \end{equation}
    achieves a long-term average AoI given by:
    \begin{equation} \label{aoi-mad-zw-age}
    \begin{aligned}
                \lambda^\text{MAD-ZW} &= \sum_{i=1}^N p_i \mu_i \prod_{k=i+1}^N (1-p_k)
        \\&+
        \frac{
        \sum_{i=1}^N 
            p_i \mu_i 
            \left(
            \frac{\mu_i}{2} + \frac{\sigma_i^2}{2\mu_i}
            \right)
            \prod_{k=i+1}^N (1-p_k)
        }{
        \sum_{i=1}^N 
            p_i \mu_i 
            \prod_{k=i+1}^N (1-p_k)
        }.
    \end{aligned}
    \end{equation}        
    \end{Lemma}
    \begin{proof}
        See Appendix~\ref{proof:lemma-mad-zw}.
    \end{proof}
    This policy can be undesirable over simpler policies like route $k$-Zero Wait. Following is an analysis where $N=3$ and $\boldsymbol{p}=[1,p,p]$.
    \paragraph{MAD-Zero Wait vs. route $1$-Zero Wait}\label{madzwvsr1zw}
    The analytical expression for $\lambda^{MAD-ZW}$ can be obtained from \eqref{aoi-mad-zw-age} by setting $N=3$ and $\boldsymbol{p}=[1,p,p]$. Then,
    \begin{equation}
        \lambda^{MAD-ZW} = A(p) + \frac{B(p)}{A(p)},
    \end{equation}
    where
    \begin{equation}
    \begin{aligned}
        &A(p) = \left(1-p^2\right)\mu_1 + \left(p-p^2\right)\mu_2 + p\mu_3\\
        &B(p) = \left(
            \frac{\mu_1}{2} + \frac{\sigma_1^2}{2\mu_1}
            \right)\left(1-p^2\right)\mu_1 \\&\quad\quad\ + \left(
            \frac{\mu_2}{2} + \frac{\sigma_2^2}{2\mu_2}
            \right) \left(p-p^2\right)\mu_2 + \left(
            \frac{\mu_3}{2} + \frac{\sigma_3^2}{2\mu_3}
            \right)p\mu_3.
    \end{aligned}   
    \end{equation}
    As a result, if the first derivative of $\lambda^{MAD-ZW}$ with respect to $p$, given by:
    \begin{equation}
        \frac{d}{dp}\lambda^{MAD-ZW} = A'(p) + \frac{B'(p)\,A(p) - B(p)\,A'(p)}{A(p)^2},
    \end{equation}
    satisfies $\frac{d}{dp}\lambda^{MAD-ZW}>0$ for $p\in[0,1]$, then the MAD-Zero Wait age will increase with $p$. In such cases, route $1$-Zero Wait policy will outperform the MAD-Zero Wait.

    \noindent $\bullet$ \textit{Single Route Routing with Zero-Wait Sampling (\textbf{\textbf{Route $X$-ZW}})}: This policy always selects route $x$. It is combined with a zero-wait strategy.

    \noindent $\bullet$ \textit{Single Route Routing with AoI-Optimal Sampling (\textbf{Route $X$ Optimal})}: This policy always selects route $x$ and follows the AoI-optimal waiting strategy as outlined in \cite[Theorem 4]{sun2017update}.
    
	\subsection{Satellite-Terrestrial Integrated Routes}
	We consider two distinct classes of routes, denoted by $\mathcal{N}_{\text{Sat}}$ and $\mathcal{N}_{\text{Ter}}$. Here, $\mathcal{N}_{\text{Sat}}$ represents the set of Low Earth Orbit (LEO) Satellite routes with stochastic delays, while $\mathcal{N}_{\text{Ter}}$ represents the set of terrestrial routes with stochastic delays. 

	\subsubsection{LEO Satellite Routes with Stochastic Delays}
	For $l \in \mathcal{N}_{\text{Sat}}$, the delay is modeled by a \textit{log-normal distribution}, characterized by the following probability density function \cite{peebles2001probability}:
	\begin{equation}
		P_{Y\sim Q_l}(y)=\frac{1}{y \beta_l \sqrt{2\pi}} \exp\left( -\frac{(\ln y - \alpha_l)^2}{2\beta_l^2} \right), l\in\mathcal{N}_{\text{Sat}},
	\end{equation}
	where $\alpha_l$ and $\beta_l$ correspond to the mean and standard deviation of the underlying normal distribution.
    
	The mean $\mu_l$ and the variance $\sigma_l^2$ of $Y\sim Q_l$ are given by:
	\begin{subequations}
		\begin{align}
			&\mu_l = \exp( \alpha_l + \frac{\beta_l^2}{2}),l\in\mathcal{N}_{\text{Sat}}\\
			&\sigma_l^2 = ( \exp(\beta_l^2) - 1) \exp(2\alpha_l + \beta_l^2), l\in\mathcal{N}_{\text{Sat}}.
		\end{align}
	\end{subequations}
    
	\subsubsection{Terrestrial Routes with Stochastic Delays} If $l\in\mathcal{N}_{\text{Ter}}$, we leverage the \textit{gamma distribution} to simulate the statistics of delay~$y$, where the probability density function is given by \cite{peebles2001probability}:
	\begin{equation}
		P_{Y\sim Q_l}(y)=\frac{1}{\Gamma(\theta_l) {\gamma_l}^{\theta_l}} y^{\theta_l - 1} e^{-y/\gamma_l},l\in\mathcal{N}_{\text{Ter}}.
	\end{equation}
    
	The mean $\mu_l$ and the variance $\sigma_l^2$ of $Y\sim Q_l$ are given by:
		\begin{align}
			\mu_l = \theta_l\gamma_l, \text{ and } \sigma_l^2 = \theta_l\gamma_l^2,l\in\mathcal{N}_{\text{Ter}}.
		\end{align}

	\subsection{Parameter Settings}

    \begin{table}
	\caption{Simulation Parameters $E_{\max}=\infty$}
	\label{tab:sim:A}
	\resizebox{\linewidth}{!}{
	\begin{tabular}{cccc}
		\toprule
		\textbf{Route} & \textbf{Route $1$} & \textbf{Route $2$} & \textbf{Route $3$} \\
		\midrule
		\textbf{Parameters} & $(Q_1,\mu_1, \sigma_1, p_1)$ & $(Q_2,\mu_2, \sigma_2, p_2)$ & $(Q_3,\mu_3, \sigma_3, p_3)$ \\
		\midrule
		Fig. \ref{figsimi1}, $E_{\max} = \infty$ & $(\text{Gamma},6,2,1)$ & $(\text{Log-normal},5,4,p)$ & $(\text{Gamma},3,7,p)$ \\
		Fig. \ref{figsimi2}, $E_{\max} = \infty$ & $(\text{Gamma},10,8,1)$ & $(\text{Log-normal},4,4,p)$ & $(\text{Log-normal},3,6,p)$ \\
		\bottomrule
	\end{tabular}
	}
\end{table}

\begin{table}[t]
\centering
\caption{Impact of exponential growth rate $\alpha$ and sensitivity to $\mu_1$ variations where $Y_{\text{cap}} = 50$. The table reports the average penalty cost for single-route optimal policies versus the optimal policy under the nominal $\mu_1$, as well as the optimal policy's performance when $\mu_1$ varies by $\pm 10\%$.}
\label{tab:alpha_sensitivity}
\resizebox{\linewidth}{!}{
\begin{tabular}{c c c c | c c | c c}
\toprule
\multirow{2}{*}{\boldmath{$\alpha$}} & \multicolumn{3}{c}{\textbf{Nominal Case}} & \multicolumn{2}{c}{\boldmath{$+10\% \mu_1$}} & \multicolumn{2}{c}{\boldmath{$-10\% \mu_1$}} \\ 
\cmidrule(lr){2-4} \cmidrule(lr){5-6} \cmidrule(l){7-8}
 & \textbf{Route 1} & \textbf{Route 2} & \textbf{Optimal} & \textbf{Route 1} & \textbf{Optimal} & \textbf{Route 1} & \textbf{Optimal} \\ 
\midrule
0.1 & 1.844 & 1.827 & \textbf{1.5748} & 1.9627 & \textbf{1.63} & 1.7555 & \textbf{1.5314}\\
0.15 & 11.88 & 11.89 & \textbf{9.34} & 14.576 & \textbf{10.313} & 9.785 & \textbf{8.2} \\
0.2 & 9.705 & 9.734 & \textbf{6.908} & 14.141 & \textbf{8.0139} & 6.7628 & \textbf{5.457}\\
0.3 & 10.026 & 10.072 & \textbf{5.812} & 20.381 & \textbf{7.2419} & 5.0575 & \textbf{3.5815}\\
1 & 11.031 & 11.299 & \textbf{4.482} & 35.716 & \textbf{7.1141} & 3.4699 & \textbf{1.8937}\\
\bottomrule
\end{tabular}%
}
\end{table}
    
\begin{table}[t]
\centering
\caption{Performance comparison under different penalty saturation values $Y_{\text{cap}}$ (with fixed $\alpha = \text{1}$). Parentheses denote the cost ratio relative to the optimal policy (Benchmark Cost/Optimal Cost).}
\label{tab:ycap_impact}
\begin{tabular}{c c c c}
\toprule
\multirow{2}{*}{\boldmath{$Y_{\text{cap}}$}} & \multicolumn{3}{c}{\textbf{Average Penalty Cost}} \\ 
\cmidrule(lr){2-4}
 & \textbf{Route 1} & \textbf{Route 2} & \textbf{Optimal} \\ 
\midrule
50 & 11.031 {\color{gray}\scriptsize($2.46\times$)} & 11.299 {\color{gray}\scriptsize($2.52\times$)} & \textbf{4.482}\\
100 & 2.705 {\color{gray}\scriptsize($2.96\times$)} & 2.694 {\color{gray}\scriptsize($2.95\times$)}& \textbf{0.9127}\\
200 & 3.244 {\color{gray}\scriptsize($3.54\times$)}& 3.224 {\color{gray}\scriptsize($3.52\times$)} & \textbf{0.9164}\\
400 & 7.976 {\color{gray}\scriptsize($4.11\times$)} & 7.987 {\color{gray}\scriptsize($4.12\times$)} & \textbf{1.939}\\
\bottomrule
\end{tabular}
\end{table}
        \begin{figure}[t]
		\centering
		\includegraphics[width=0.95\linewidth]{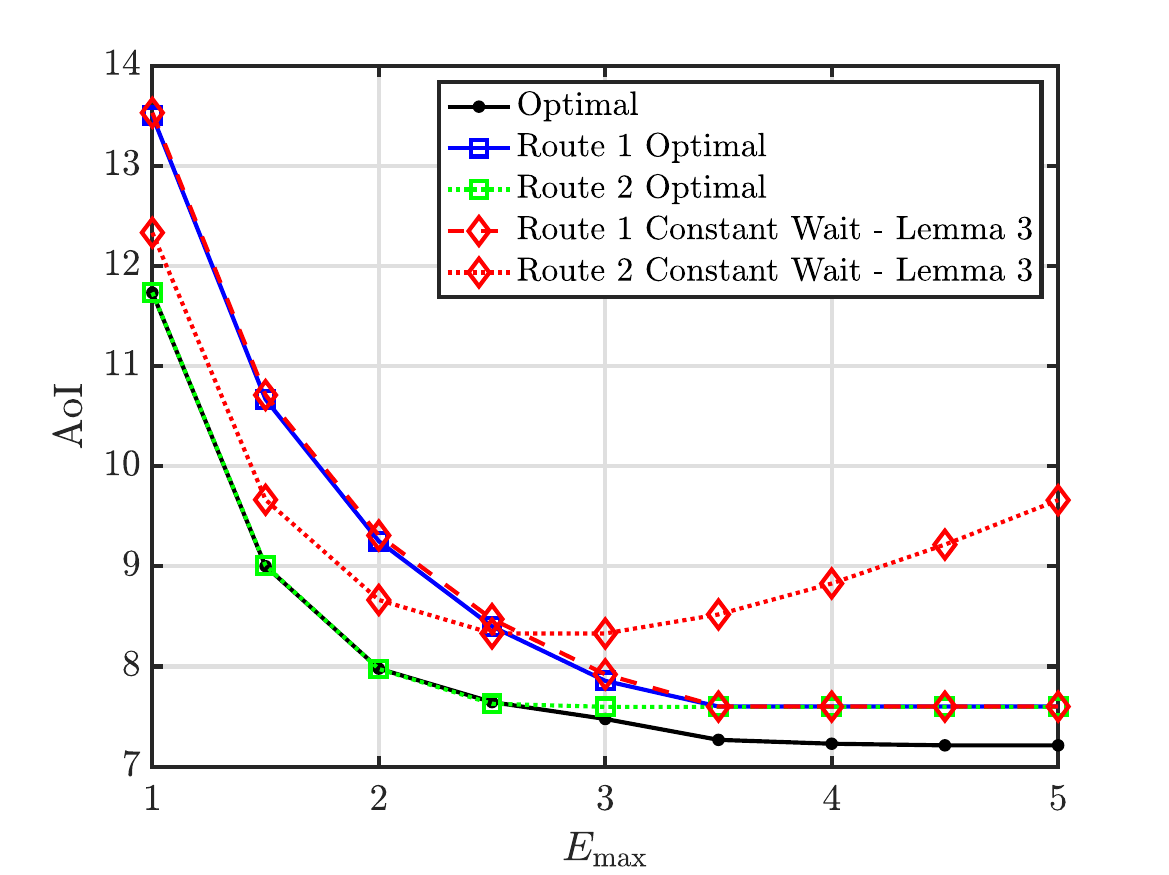} 
		\caption{AoI vs. $E_{\max}$}
		\label{figsimenergy}
	\end{figure}
        \begin{figure}[t]
		\centering
		\includegraphics[width=0.95\linewidth]{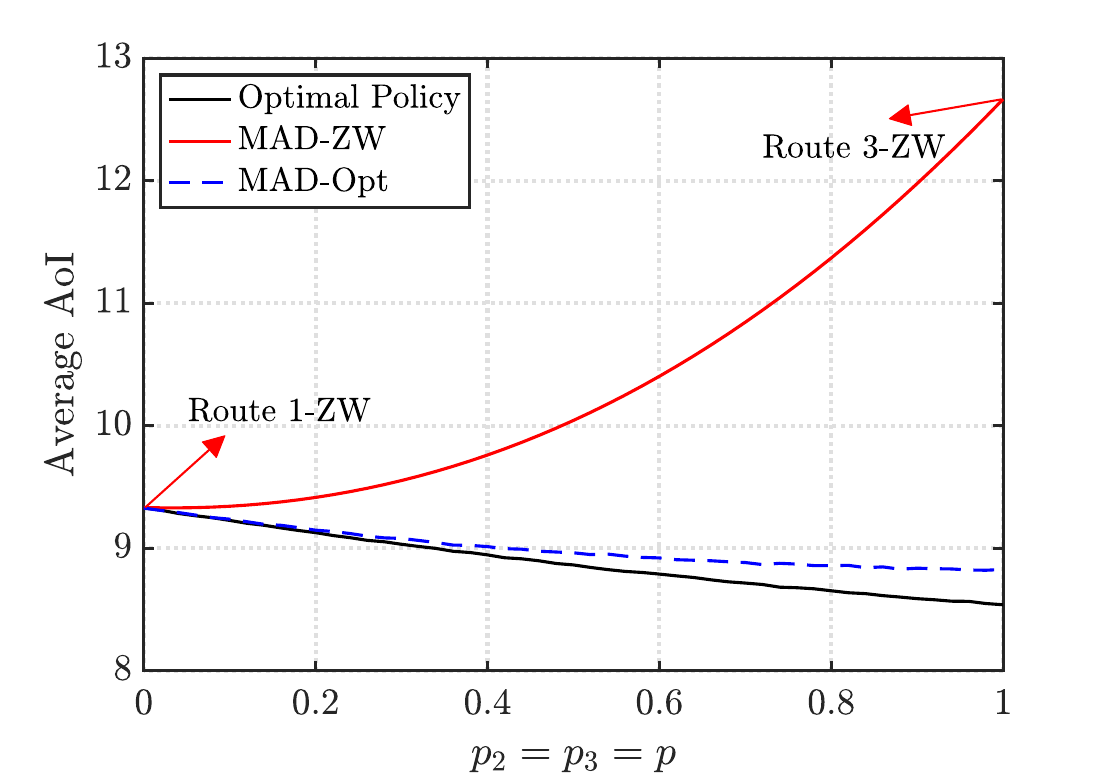} 
		\caption{Simulation results of systems with $N=3$ and competitive route $1$.}
		\label{figsimi1}
	\end{figure}
    \begin{figure}[t]
		\centering
		\includegraphics[width=0.95\linewidth]{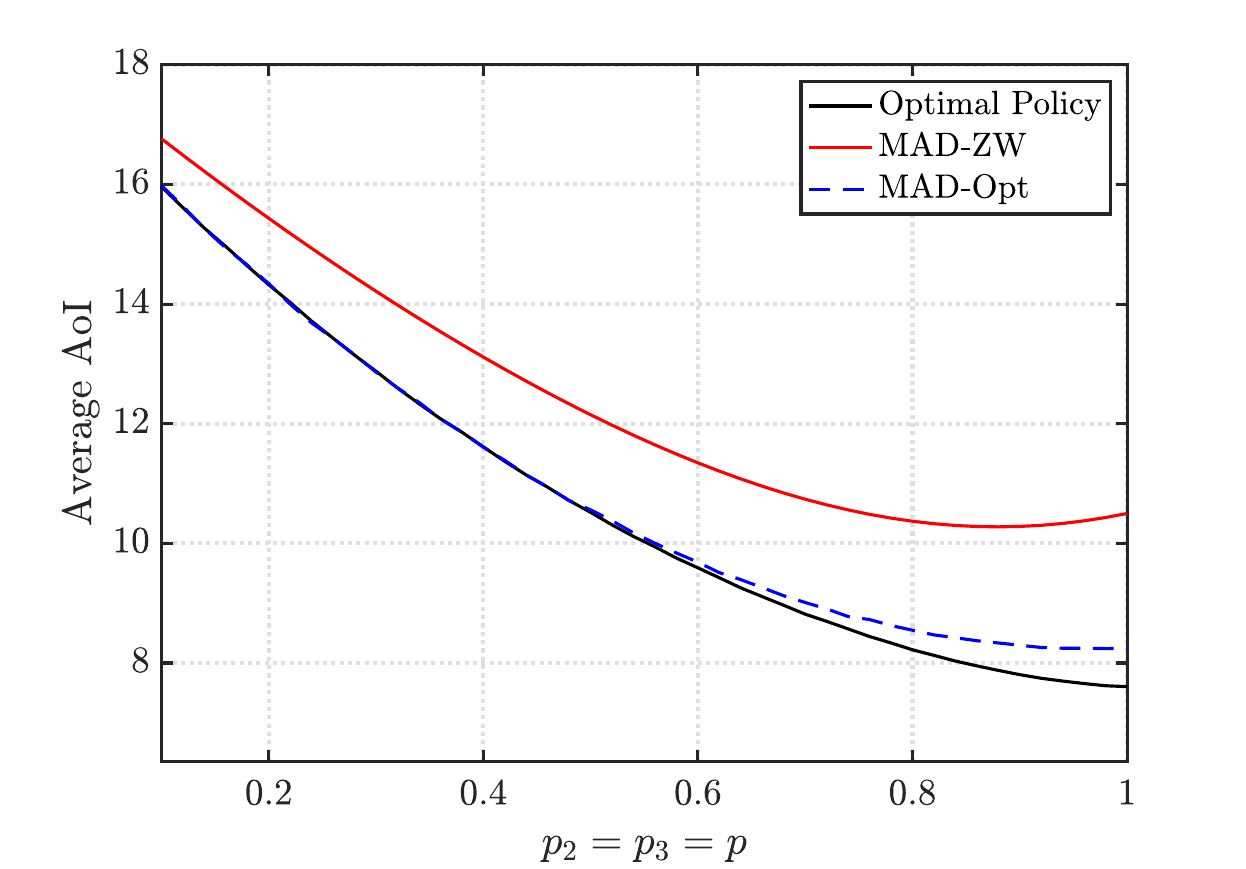} 
		\caption{Simulation results of systems with $N=3$ and uncompetitive route $1$.}
		\label{figsimi2}
	\end{figure}
    
    To evaluate the performance of the proposed policy under non-linear aging costs, we adopt a truncated exponential penalty function defined as:
    \begin{equation}
        f(\Delta) = a \left( \min\left( e^{\alpha \Delta}, e^{\alpha Y_{\text{cap}}} \right) - 1 \right),
    \end{equation}
    where $\alpha > 0$ determines the growth rate of the penalty, $Y_{\text{cap}}$ imposes a maximum saturation level on the cost, and $a$ is a variable scaling parameter used for normalization. We introduce the saturation threshold $Y_{\text{cap}}$ to reflect the practical reality that in many real-time systems, information effectively loses all utility beyond a certain age. Once the data staleness exceeds this critical level, any further delay does not alter the value of the penalty function.

    We consider a two-route system designed to represent an ``ultra-competitive'' scenario where routing decisions are non-trivial. The parameters for the second route are fixed at $\mu_2 = 0.1$ and $\sigma_2 = 6.05$ with Gamma distribution, and the standard deviation of the delay of the first route is fixed at $\sigma_1 = 3.7$ with Log-normal distribution. For each distinct value of $\alpha$ tested, the mean delay of the first route, $\mu_1$, is tuned specifically to create a scenario where both routes are statistically competitive (i.e., offering similar average age penalties). We do not consider availability or energy restrictions in this scenario.

    Finally, since $\mu_1$ is selected to create a high-gain scenario, we vary this parameter by $\pm 10\%$ to assess the stability of the result. This sensitivity analysis serves to verify that the reported gains are robust features of the proposed policy under the given penalty function, rather than fragile outcomes that vanish with minor deviations in system parameters.
    
    Secondly, we consider a scenario where there are $2$ available routes that are always available with the varied energy constraint $E_{\max} \in [1,5]$, under the linear age penalty function $f(\Delta)=\Delta$. The routes have the delay statistics: Log-normal distribution, $\mu_1 = 5, \sigma_1 = 1$; Gamma distribution, $\mu_2 = 1, \sigma_2 = 7.3$. Route dependent transmission costs are given by: $G_1 = 3,\; G_2 = 18 $, and the sampling cost is $C_s = 2$.

    We then consider a scenario with no energy constraint with three available routes where $\mathcal{N} = \{1, 2, 3\}$. 
The parameter setting for the simulations is presented in Table ~\ref{tab:sim:A} where $p_2 = p_3 = p$ is the varying parameter.
        
	\subsection{Discussions}

    \subsubsection{Impact of Penalty Parameters of Non-Linear Penalty Functions}

    Table \ref{tab:alpha_sensitivity} illustrates the system performance as the exponential growth rate $\alpha$ varies. The parameter $\alpha$ essentially dictates the system's sensitivity to aging; a larger $\alpha$ imposes a steeper penalty on outdated information, thereby placing a higher premium on maintaining a low AoI. We observe that the relative benefit of the proposed joint policy becomes more pronounced in regimes with higher $\alpha$. As the penalty for high age explodes, intelligent routing becomes critical. Specifically, for $\alpha=1$, the optimal policy achieves a penalty reduction of nearly $60\%$ compared to the best single-route alternative. This confirms that in systems where stale information is costly, dynamic routing yields substantial gains over static strategies.

    Furthermore, the sensitivity analysis with $\mu_1 \pm 10\%$ demonstrates the robustness of these gains. The improvements persist even when the delay statistics deviate from the nominal "ultra-competitive" configuration, confirming that the policy's advantage is not an artifact of highly specific parameter tuning. A striking example is observed at $\alpha=1$ when Route 1 degrades ($\mu_1 + 10\%$). Although the stand-alone cost of Route 1 is more than three times that of Route 2, the optimal policy does not simply discard it. Instead, by opportunistically utilizing Route 1, the joint policy achieves a cost reduction of over $37\%$ compared to relying solely on the "better" Route 2. This highlights the policy's ability to extract value from diversity, even when route qualities are highly asymmetric.

    Table \ref{tab:ycap_impact} examines the effect of the saturation threshold $Y_{\text{cap}}$. This parameter represents the "uselessness threshold" of the information; beyond this age, the data is considered equally stale, and the penalty remains constant. The results indicate a clear trend: increasing $Y_{\text{cap}}$ amplifies the gains achieved by the optimal policy. When the cap is low, the penalty function is flatter, dampening the impact of poor routing decisions. However, as $Y_{\text{cap}}$ increases, the system is exposed to the full cost of aging for longer durations. Notably, for $Y_{\text{cap}} = 400$, the optimal policy reduces the average penalty by over $75\%$ compared to single-route baselines.

    \subsubsection{Impact of Energy Constraints}

    Fig.~\ref{figsimenergy} illustrates the impact of the energy budget $E_{\max}$ on system performance. As expected, the long-term average AoI decreases as the energy constraint is relaxed, eventually converging to a constant floor once $E_{\max}$ exceeds the consumption required for the unconstrained optimum.

    A noteworthy observation is that the energy consumption of the unconstrained joint optimal policy ($E = 4.14$) exceeds that of the unconstrained policies for individual routes ($E_1 = 3.4$, $E_2 = 2.75$). This increase stems from the threshold-based routing structure: the joint policy avoids Route~2 in low-delay regions, effectively reducing the average waiting time required when the route \emph{is} selected. This reduction in waiting time leads to a higher transmission frequency, thereby increasing the overall energy expenditure of the system compared to the single-route scenarios.

    We consider two routes that yield approximately the same optimal AoI in the unconstrained regime. A striking outcome is that, despite incurring six times higher transmission cost ($G_2 = 6 G_1$), Route~2 becomes the \emph{exclusive} choice of the optimal policy when $E_{\max}$ is small. This behavior arises from the distinct delay characteristics of the routes. Route~1, characterized by low delay variance, relies on frequent updates (approaching zero-wait) to minimize age, leading to high average energy consumption. Conversely, Route~2's high delay variance necessitates large waiting times to mitigate age penalties; these extended idle periods naturally reduce the average transmission frequency. Consequently, under strict energy constraints, the system favors Route~2 as it sustains operation at a lower average power draw. 

    It is also important to note that in this low-energy regime, the dynamic routing policy offers no discernible benefit over the best single-route strategy. This indicates that when energy is severely limited, the optimal strategy simplifies to the static selection of the single most viable route, as the system lacks the energy budget to exploit diversity gains.

    \subsubsection{Impact of Route Availability}

    Fig.~\ref{figsimi1} demonstrates that MAD-Zero Wait can underperform against a simpler policy. When $p=0$, MAD-Zero Wait performs similar to the optimal policy since route $1$'s delay has a small variance in this setting. As $p$ increases, the MAD-Zero Wait policy uses the other routes, which are not suitable to zero wait policies, increasingly often. When $p=1$, MAD-Zero Wait policy is equivalent to the route $3$-Zero Wait policy. An analysis showing when we can expect MAD-Zero Wait to exhibit this behavior was given in subsection~\ref{benchmark}.

    Figures~\ref{figsimi1} and \ref{figsimi2} demonstrate that the advantage of using the optimal policy over the MAD-Optimal policy dwindles in the low availability region (small $p$). It can also be seen that the separation between the policies happens for a larger $p$ when route $1$ has worse delay characteristics. As a result, it may be beneficial to apply the proposed MAD-Optimal policy (smaller complexity) in cases where the routes in $\mathcal{R}_{\infty}$ are uncompetitive and routes in $\mathcal{R}_{<\infty}$ have low joint availability (i.e, when $\big(\prod_{i \in \mathcal{R}_{<\infty}} 1-p_i\big)$ is large).

\section{Conclusion}
In this work, we investigated a multi-route status update system and proved that a threshold-based joint sampling and routing policy can minimize the long-term average AoI. We introduced an efficient algorithm namely Bisec-\textsc{ReaVI} to compute this optimal policy. Our simulations consistently show improvements in AoI, revealing that higher variance or mean delays in certain routes can still help minimize AoI when jointly optimized. This challenges the common intuition that lower delay variance always leads to better AoI performance and provides insights into routing design for future TN-NTN.

\bibliographystyle{IEEEtran}
\bibliography{references}

\appendices

\section{Proof of Lemma \ref{lem1}}\label{appa}

\textbf{Part 1}. 
We first prove that
\begin{equation}\label{eq456}
\lambda_c^\star \le \lambda
\Longleftrightarrow
h(\lambda,c) \le 0.
\end{equation}
If $\lambda_c^\star \le \lambda$, by the definition of the optimal Lagrangian value, there exists a policy $\pi$ such that:
\begin{equation} \label{eqg1}
\begin{aligned}
&\limsup_{n\to\infty}\\&\frac{\sum_{i=0}^{n-1} \mathbb{E}_{\pi}\big[q(Y_i, Z_i, Y_{i+1}) + c(E_{i+1} - E_{\max}(Z_i+Y_{i+1}))\big]}
{\sum_{i=0}^{n-1} \mathbb{E}_{\pi}[Z_i + Y_{i+1}]} \\&\le \lambda. 
\end{aligned}
\end{equation}
Moving $\lambda$ to the left-hand side and combining terms over the common denominator yields: $\exists~\pi$ such that

\begin{equation} \label{eqg2}
\limsup_{n\to\infty}\frac{\frac{1}{n}
\sum_{i=0}^{n-1}
\mathbb{E}_{\pi}[\Phi_i]}{\frac{1}{n}\sum_{i=0}^{n-1} \mathbb{E}_{\pi}[Z_i + Y_{i+1}]}
\le
0,
\end{equation}
where
\begin{equation}
\begin{aligned}
\Phi_i \triangleq & \; q(Y_i, Z_i, Y_{i+1}) - \lambda(Z_i + Y_{i+1}) \\ 
&- c\big(E_{\max}(Z_i+Y_{i+1}) - E_{i+1}\big).
\end{aligned}
\end{equation}
Since $Y_i$'s over the same route are independent, the inter-sampling times $T_i = Y_i + Z_i$ are regenerative. Since there are $N$ routes, the expected period of the most frequently used route satisfies $\mathbb{E}[n_{k+1}-n_k]\le N$, where $n_k$ denotes the $k$-th epoch a particular route is used. Because $T_i$'s are regenerative and we have $0<\mathbb{E}[D_{n_{k+1}}-D_{n_k}]<\infty$, for all $k$, the renewal theory \cite{ross1995stochastic} tells us that $\lim_{n\to\infty}\frac{1}{n}\sum_{i=0}^{n-1} \mathbb{E}[Z_i + Y_{i+1}]$ exists and is positive. Thus, there exists a policy $\pi$ such that the numerator of the left-hand side \eqref{eqg2} is less than zero, which indicates that the infimum of the numerator in \eqref{eqg2} is less than zero, indicating that $h(\lambda,c)\le0$.

Conversely, if $h(\lambda,c) \le 0$, as $\lim_{n\to\infty}\frac{1}{n}\sum_{i=0}^{n-1} \mathbb{E}[Z_i + Y_{i+1}]$ exists and is positive, we can derive \eqref{eqg1} and \eqref{eqg2}, which indicates that  $\lambda_c^\star\le\lambda$. 
The corollary $\lambda_c^\star > \lambda
\Longleftrightarrow
h(\lambda,c) > 0$ can be derived directly from \eqref{eq456} by leveraging \textit{Modus Tollens}.

\noindent\textbf{Part 2:} $\lambda_c^\star = \lambda
\Longleftrightarrow
h(\lambda,c) = 0$. 
If $h(\lambda,c)=0$, from part 1 of the proof, we can first establish that $\lambda_c^\star\le\lambda$. We then show that the policy $\pi$ such that $h(\lambda,c)=0$ can lead to
\begin{equation}\label{lambda}
\begin{aligned}
&\limsup_{n\to\infty}\\&\frac{\sum_{i=0}^{n-1} \mathbb{E}_{\pi}\big[q(Y_i, Z_i, Y_{i+1}) + c(E_{i+1} - E_{\max}(Z_i+Y_{i+1}))\big]}
{\sum_{i=0}^{n-1} \mathbb{E}_{\pi}[Z_i + Y_{i+1}]} \\&= \lambda,
\end{aligned}
\end{equation}
which indicates that $\lambda\ge\lambda_c^\star$. Combining these together, we can obtain $\lambda=\lambda_c^\star$. Conversely, if $\lambda=\lambda_c^\star$, we can establish from Part 1 that $h(\lambda,c)\le0$. Meanwhile the definition of $\lambda^\star$ in \eqref{eq1-} leads to
\begin{equation}\label{lambda2}
\begin{aligned}
\forall~\pi,~\limsup_{n\to\infty}\frac{1}{n}\sum_{i=0}^{n-1} \mathbb{E}_{\pi}[\Phi_i] \ge 0,
\end{aligned}
\end{equation}
which indicates that $h(\lambda,c)\ge0$. Combining these together, we establish that $h(\lambda,c)=0$.

\section{Proof of Lemma \ref{lem2}}\label{prooflemma5}
We start with the general form of the action-value function $Q(y,r)$. In the non-linear setting, the accumulated age penalty is captured by the primitive function $F(t) \triangleq \int_0^t f(\tau) d\tau$. The simplified action-value function, substituting the optimal waiting policy $z^\star = (\beta^\star(r) - y)^+$, is given by:
\begin{equation}\label{Q_nonlinear_full}
    \begin{aligned}
        Q(y,r) = \;& C_r - (\lambda^\star + c E_{\max}) (\beta^\star(r) - y)^+ \\
        & + \mathbb{E}_{Y'\sim Q_r}\left[ F(y + (\beta^\star(r) - y)^+ + Y') - F(y) \right],
    \end{aligned}
\end{equation}
where $C_r$ collects terms independent of $y$ (such as the value function expectation and energy costs). We analyze the partial derivative $\frac{\partial Q(y,r)}{\partial y}$ in two regions defined by the threshold $\beta^\star(r)$.

\textbf{Case 1: Waiting Region ($y < \beta^\star(r)$).}
In this region, the waiting time is positive: $z^\star = \beta^\star(r) - y$.
Substituting this into \eqref{Q_nonlinear_full}, the linear term becomes $-(\lambda^\star + c E_{\max})(\beta^\star(r) - y)$.
The argument of the penalty function becomes $y + (\beta^\star(r) - y) + Y' = \beta^\star(r) + Y'$, which is constant with respect to $y$.
Thus, the equation simplifies to:
\begin{equation}
\begin{aligned}
      Q(y,r) &= C_r - (\lambda^\star + c E_{\max})(\beta^\star(r) - y) \\&+ \mathbb{E}_{Y'}[F(\beta^\star(r) + Y') - F(y)].  
\end{aligned}
\end{equation}
Differentiating with respect to $y$:
\begin{equation}
    \frac{\partial Q(y,r)}{\partial y} = (\lambda^\star + c E_{\max}) - f(y).
\end{equation}
Recall the definition of the threshold $\beta^\star(r) = \inf \{ t \ge 0 : \mathbb{E}[f(t + Y')] \ge \lambda^\star + c E_{\max} \}$.
Since $y < \beta^\star(r)$, $y$ is strictly smaller than the infimum. This implies that $y$ does not satisfy the condition of the set. Therefore, the expected penalty at $y$ must be strictly less than the system cost:
\begin{equation}
    \mathbb{E}[f(y + Y')] < \lambda^\star + c E_{\max}.
\end{equation}
Furthermore, since $Y' \ge 0$ and $f(\cdot)$ is monotonically non-decreasing, we have $f(y) \le \mathbb{E}[f(y + Y')]$.
Combining these inequalities yields:
\begin{equation}
    f(y) \le \mathbb{E}[f(y + Y')] < \lambda^\star + c E_{\max}.
\end{equation}
Consequently, $(\lambda^\star + c E_{\max}) - f(y) > 0$, proving the derivative is strictly positive in this region.

\textbf{Case 2: Sampling Region ($y \ge \beta^\star(r)$).}
In this region, the waiting time is zero: $z^\star = 0$.
The linear term vanishes, and the equation simplifies to:
\begin{equation}
    Q(y,r) = C_r + \mathbb{E}_{Y'}[F(y + Y') - F(y)].
\end{equation}
Differentiating with respect to $y$:
\begin{equation}
    \frac{\partial Q(y,r)}{\partial y} = \mathbb{E}_{Y'}[f(y + Y')] - f(y).
\end{equation}
Since $Y' \ge 0$ and $f(\cdot)$ is a monotonic non-decreasing function, it holds that $\mathbb{E}_{Y'}[f(y + Y')] \ge f(y)$. Thus, the derivative is always non-negative.

\textbf{Conclusion:}
Combining both cases, we have $\frac{\partial Q(y,r)}{\partial y} \ge 0$ for all $y \in \mathbb{R}^+$. This proves statement (1). Since $V^*(y,\mathbf{l}) = \min_{r \in \mathcal{R}(\mathbf{l})} Q(y,r)$ is the minimum of monotonically increasing functions, it is also monotonically increasing with $y$, proving statement (2).

\section{Proof of Lemma \ref{lem3}}\label{prooflemma6}

The target threshold is defined as $\beta^\star(r) = \inf \{ t \ge 0 : \Psi_r(t) \ge \lambda^\star + cE_{\max} \}$.

First, we determine the ordering of the thresholds. Consider any $t$ that satisfies the threshold condition for route $k$, i.e., $\Psi_k(t) \ge \lambda^\star + cE_{\max}$. By the ordering condition \eqref{eq:ordering_condition}, we have $\Psi_j(t) \ge \Psi_k(t)$ for all $t$. Therefore, inequality $\Psi_j(t) \ge \lambda^\star + cE_{\max}$ holds.
This implies that any time $t$ feasible for the infimum of route $k$ is also feasible for route $j$. Since the set of feasible times for route $k$ is a subset of that for route $j$, the infimum of the larger set must be smaller or equal:
\begin{equation}
    \beta^\star(j) \le \beta^\star(k).
\end{equation}

Next, we evaluate the partial derivative of the action-value function with respect to $y$. From Appendix~\ref{prooflemma5}, the derivative is given by:
\begin{equation}
    \frac{\partial Q(y,r)}{\partial y} = 
    \begin{cases} 
        \lambda^\star + cE_{\max} - f(y), & y < \beta^\star(r) \\ 
        \Psi_r(y) - f(y), & y \geq \beta^\star(r)
    \end{cases}
\end{equation}
We analyze the difference $\Delta_{jk}(y) \triangleq \frac{\partial Q(y,j)}{\partial y} - \frac{\partial Q(y,k)}{\partial y}$ across the three intervals defined by $\beta^\star(j) \le \beta^\star(k)$:

\begin{itemize}
    \item \textbf{Region 1} ($y < \beta^\star(j)$): Both routes are in the waiting region.
    \begin{equation}
        \Delta_{jk}(y) = (\lambda^\star + cE_{\max} - f(y)) - (\lambda^\star + cE_{\max} - f(y)) = 0.
    \end{equation}
    
    \item \textbf{Region 2} ($\beta^\star(j) \le y < \beta^\star(k)$): Route $j$ is in the sampling region, while route $k$ is in the waiting region.
    \begin{equation}
        \begin{aligned}
        \Delta_{jk}(y) &= (\Psi_j(y) - f(y)) - (\lambda^\star + cE_{\max} - f(y)) \\
        &= \Psi_j(y) - (\lambda^\star + cE_{\max}).
        \end{aligned}
    \end{equation}
    Since $y \ge \beta^\star(j)$ and $\Psi_j(y)$ is non-decreasing, by the definition of the infimum, we must have $\Psi_j(y) \ge \lambda^\star + cE_{\max}$. Thus, $\Delta_{jk}(y) \ge 0$.

    \item \textbf{Region 3} ($y \ge \beta^\star(k)$): Both routes are in the sampling region.
    \begin{equation}
        \begin{aligned}
        \Delta_{jk}(y) &= (\Psi_j(y) - f(y)) - (\Psi_k(y) - f(y)) \\
        &= \Psi_j(y) - \Psi_k(y).
        \end{aligned}
    \end{equation}
    By the assumed ordering condition \eqref{eq:ordering_condition}, $\Psi_j(y) \ge \Psi_k(y)$ holds for all $y$, so $\Delta_{jk}(y) \ge 0$.
\end{itemize}

Combining these cases, we conclude that $\frac{\partial Q(y,j)}{\partial y} \geq \frac{\partial Q(y,k)}{\partial y}$ for all $y \in \mathbb{R}^+$.

            \section{Proof of Lemma \ref{cor1}}\label{lemma6:proof}
            Since route $j$ is optimal at $y = y^\star, \mathcal{R}$, we have
					\begin{equation} \label{eqcor1}
						Q(y^\star,j) \le Q(y^\star,i),
					\end{equation}for any $i \in \mathcal{R}$.
					Now, for $i<j$ we know $\mu_i \ge \mu_j$. Then, combining \eqref{lem3eq} with \eqref{eqcor1} we obtain
					\begin{equation}
						Q(y,j) \le Q(y,i), ~ y\ge y^\star,\mathcal{R},
					\end{equation}which proves that no route $i<j, \;i \in A$ can be optimal for $y > y^\star, \mathcal{R}$. The proof for the converse statement follows the same logic.
                    
\section{Proof of Lemma \ref{lem_thrbound}}\label{proof:lemma1}

We know from Lemma \ref{cor1} that the optimal route $j$ at $y = \tau_k(\mathbf{l}_i)$ satisfies $\mu_j < \mu_{a_k(\mathbf{l}_i)}$. 
Then, we have \begin{equation} 
\frac{\partial Q(y,j)}{\partial y} = \frac{\partial Q\big(y,\mathbf{l}_k(\mathbf{l}_i)\big)}{\partial y} = \lambda^\star -y,\quad y < \beta_k^\star(\mathbf{l}_i).
\end{equation}
Therefore, $\tau_k(\mathbf{l}_i)$ must be greater than $\beta_k^\star(\mathbf{l}_i)$.

\section{Proof of Lemma \ref{lembounds}} \label{ap3.2.4}
Consider a stationary policy $\pi = (k,w_k,k,w_k,\ldots)$ that consistently selects a single route $k \in \mathcal{R}_{\infty}$ (where $\mathcal{R}_{\infty}$ denotes the set of routes that are always available) and waits for a constant time $w_k$ before generating a new sample after the previous update is delivered.

The average energy expenditure of this policy is given by
\begin{equation}
    \frac{C_s + G_k\mu_k}{\mu_k + w_k} \le \frac{C_s + G_k\mu_k}{\frac{C_s + G_k\mu_k}{E_{\max}}} = E_{\max}.
\end{equation}
Thus, $\pi$ is a valid policy.

Next, we derive the long-term average age penalty $\lambda_k^{cw}$ achieved by this policy using the Renewal Reward Theorem. A renewal cycle $i$ begins immediately after the delivery of packet $i-1$. The age is reset to $Y_{i-1}$ (the delay of the previous packet). The system waits for $w_k$, and then transmits packet $i$, which experiences delay $Y_i$. The cycle ends when packet $i$ is delivered at time $t = w_k + Y_i$.

The accumulated penalty during one cycle is the integral of the penalty function $f(\Delta(t))$. The instantaneous age starts at $\Delta(0) = Y_{i-1}$ and grows linearly to $Y_{i-1} + w_k + Y_i$ at the moment of delivery. Thus, the total penalty area for cycle $i$ is:
\begin{equation}
    R_i = \int_{0}^{w_k + Y_i} f(Y_{i-1} + t) \, dt.
\end{equation}
Using the substitution $u = Y_{i-1} + t$ and the primitive function $F(t) \triangleq \int_0^t f(\tau) d\tau$, this integral becomes:
\begin{equation}
    R_i = F(Y_{i-1} + w_k + Y_i) - F(Y_{i-1}).
\end{equation}
Taking the expectation over the i.i.d. delays $Y_{i-1}, Y_i \sim Q_k$, the average penalty is:
\begin{equation}
    \lambda_k^{cw} = \frac{\mathbb{E}_{Y, Y' \sim Q_k} [ F(Y + w_k + Y') - F(Y) ]}{\mu_k + w_k}.
\end{equation}
By the definition of optimality, $\lambda^\star \leq \lambda^{\pi}$ for any valid policy $\pi$. Therefore:
\begin{equation}
\begin{aligned}
        &\lambda^\star \leq \min_{k\in \mathcal{R}_{\infty}} \lambda_k^{cw} \\&= \min_{k\in \mathcal{R}_{\infty}} \left( \frac{\mathbb{E}_{Y, Y' \sim Q_k} [ F(Y' + w_k + Y) - F(Y') ]}{\mu_k + w_k} \right).
\end{aligned}
\end{equation}
Finally, for the lower bound, since the age $\Delta(t) \ge 0$ and $f(\cdot)$ is monotonic, the instantaneous penalty is lower-bounded by $f(0)$. Consequently, the long-term average penalty satisfies $\lambda^\star \ge f(0)$. This completes the proof.

\section{Proof of Lemma \ref{lem-mad-zw-age}}\label{proof:lemma-mad-zw}
Let us derive the long-term average age achieved by using route $j$ under the MAD-ZW policy. The expected age over such intervals can be given by:
\begin{equation} \label{eq_lambda_j_av}
    \lambda_j^{av} = \frac{E \left [ YY' + \frac{(Y')^2}{2} \right]}{E[Y']} = E[Y] + \frac{E[Y']}{2} + \frac{var(Y')}{2E[Y']},
\end{equation}
where $Y$ denotes the randomized previous delay and $Y'$ denotes the delay over route $j$. Since the availability of routes are i.i.d over intervals, so is the route selection process of the MAD-ZW policy. The probability that route $k$ is used under $\pi^{MAD-ZW}$ is the probability of the event that route $k$ is ON and no route $l>k, l\le N$ is ON, i.e $k = \max \mathcal{R}$. Therefore, we can express the $E[Y]$ term in \eqref{eq_lambda_j_av} as:
\begin{equation} \label{eq_exp_y_mad-zw}
    E[Y] = \sum_{i=1}^N p_i \mu_i \prod_{k=i+1}^N (1-p_k).
\end{equation}
Then, combining \eqref{eq_lambda_j_av} and \eqref{eq_exp_y_mad-zw}, the average age attained over route $j$ can be expressed as:
\begin{equation}
    \lambda_j^{av} = \sum_{i=1}^N p_i \mu_i \prod_{k=i+1}^N (1-p_k) +  \left(
            \frac{\mu_j}{2} + \frac{\sigma_j^2}{2\mu_j}
            \right).
\end{equation}
Finally, since we can express the average age attained over any route, we take the time average over all $j$:
\begin{equation}
    \lambda^{MAD-ZW} = \sum_{j=1}^N \frac{ 
            p_j \mu_j \lambda_j^{av}
            \prod_{k=j+1}^N (1-p_k)
        }{
        \sum_{i=1}^N 
            p_i \mu_i 
            \prod_{k=i+1}^N (1-p_k)
        },
\end{equation}
which gives \eqref{aoi-mad-zw-age}.

\end{document}